\documentclass[11pt]{article}
\usepackage{colortbl}
\usepackage[dvipsnames]{xcolor}
\RequirePackage{amsthm,amsmath,amsfonts,amssymb}
\usepackage[authoryear]{natbib}
\RequirePackage[colorlinks,citecolor=blue,urlcolor=blue]{hyperref}
\RequirePackage{graphicx}
\usepackage{url}
\usepackage{siunitx}
\usepackage{color}
\usepackage{multirow}
\usepackage{array}
\usepackage{bbm}
\usepackage{float}
\usepackage{booktabs}
\usepackage{enumitem}
\usepackage{chngcntr}
\usepackage[total={170mm,230mm},
 left=25mm,
 top=20mm]{geometry}
\newgeometry{top=5mm,left=25mm,right=25mm,bottom=25mm} 

\def\PMp{$\text{PM}_{2.5}$}

\def\dGPD{$\delta$-GPD}
\newcommand{\rev}[1]{\textcolor{black}{#1}}

\title{\rev{Spatio-temporal fusion of reanalysis and \emph{in situ} data for censored threshold exceedances of \PMp{}}}

\author{M. Daniela Cuba, Craig Wilkie, Marian Scott, and Daniela Castro-Camilo*\\ \\\emph{School of Mathematics and Statistics, University of Glasgow}}

\date{\today}

\begin{document}
\maketitle

\baselineskip=16pt
\begin{center}

{\large{\bf Abstract}}
\end{center}
Data fusion models are widely used in air quality monitoring to integrate \emph{in situ} and large-scale gridded products, offering spatially complete and temporally detailed estimates. 
However, traditional Gaussian-based models often underestimate extreme pollution values, leading to biased risk assessments. 
To address this, we present a Bayesian hierarchical data fusion framework rooted in extreme value theory, using the Dirac-delta generalised Pareto distribution to jointly account for threshold and non-threshold exceedances while \rev{preserving the timing of exceedance and non-exceedance episodes}. 
Our model is used to describe and predict censored threshold exceedances of \PMp{} pollution in the Greater London region by using \rev{CAMS atmospheric composition reanalysis,} and \emph{in situ} observation stations from the automatic urban and rural network (AURN) run by the UK government. 
\rev{Key features of our approach} include combining data with varying spatio-temporal resolutions and fully accounting for parameter uncertainties. 
Results show that our model outperforms Gaussian-based alternatives and standalone \rev{reanalysis} data in predicting threshold exceedances at the majority of observation sites and can even result in improved spatial patterns of \PMp{} pollution than those discernible from the \rev{background} data.
Moreover, our approach captures greater variability and spatial patterns, such as higher \PMp{} concentrations near coastal areas, which are not evident in \rev{the reanalysis data} alone.

\par\vfill\noindent

\restoregeometry 
\newgeometry{top=20mm,left=25mm,right=20mm,bottom=20mm}

\newpage

\section{Introduction}\label{sec:Intro}

Data fusion is the process of integrating multiple data sources to improve the representation of the phenomenon of interest. It can play a vital role in applications where \emph{in situ} measurement stations or sensors are limited spatially but earth observation data are plentiful, such as air quality monitoring. 
\rev{In this paper, our aim is to develop a data fusion framework that produces a complete spatio-temporal representation of \PMp{} threshold exceedances, combining the spatial coverage of gridded reanalysis data with the local accuracy of \emph{in situ} observations to accurately characterise extremal behaviour.}

Air pollution risk is influenced by pollutant characteristics such as concentration, size, structure, and chemical composition \citep{poschl_atmospheric_2005}. 
{Most pollution is} anthropogenic, stemming from industrial activities, fossil fuel combustion, and excessive fertiliser use \citep{kampa_human_2008}. 
Key pollutants include particulate matter, with \PMp{} ($<$\SI{2.5}{\micro\metre}) being especially harmful due to its ability to penetrate deep into the lungs, causing conditions like COPD, asthma, and lung cancer \citep{kyung2020particulate}. 
Extreme \PMp{} events significantly increase hospital admissions for cardiovascular and respiratory issues \citep{anderson_clearing_2012, zhang_risk_2021}, prompting policies to mitigate their impact. 
The WHO recommends limiting 24-hour \PMp{} exposure to a maximum average of \SI{15}{\micro\metre}$^3$ for three days a year \citep{who_who_2021}, while UK guidelines cap it at \SI{20}{\micro\metre}$^3$ (Environmental Protection England; \url{https://www.legislation.gov.uk/uksi/2023/91/made}).

Maintaining air quality guidelines (AQGs) requires efficient monitoring, but identifying and predicting extreme pollution events is hindered by limited data availability \citep{martenies_health_2015}.
In the UK, the Automatic Urban and Rural Network (AURN) has only 171 stations, resulting in low national spatial coverage and increased uncertainty in areas far from observation points. 
Alternative data sources like the Copernicus Atmosphere Monitoring Service (CAMS) managed by the European Centre for Medium-Range Weather Forecasts (ECMWF) offer global coverage and long-term records but often underestimate extreme values, smooth \PMp{} concentrations in time and space \citep{palharini_assessment_2020,stahl_why_2024}, \rev{and fail to capture local nuances observed at \emph{in situ} stations.}
The limitations of sparse \emph{in situ} monitoring and the underestimation of extremes by these products highlight the need for data fusion approaches to improve spatial coverage and capture local variations in air pollution.

{Various data fusion approaches have been developed, including geostatistical models based on kriging \citep{ferreira2000air, kunzli_ambient_2005, beauchamp_dealing_2017, beauchamp_necessary_2018, xie_review_2017}. 
However, kriging and similar Gaussian-based models tend to smooth extreme values, potentially leading to inaccurate exposure and risk assessments \citep{gressent_data_2020}. Alternative methods still often rely on Gaussian assumptions \citep{fuentes_model_2005, bogaert_bayesian_2007, banerjee_hierarchical_2015, wilkie_data_2015, gengler_integrating_2016, villejo_data_2023}. 
A notable advancement by \cite{wilkie_nonparametric_2019} extended the spatially-varying parameter framework of \cite{gelfand_spatial_2003} and \cite{berrocal_spatio-temporal_2010} by allowing model coefficients to flexibly vary in space and time, using basis functions for temporal trends and an exponential decay term for spatial dependence. 
Their Bayesian approach, built on Gaussian likelihoods with conjugate priors, enables efficient sampling from closed-form posterior distributions.}

\rev{Related work in bias correction and statistical downscaling aims to translate coarse-resolution model output into locally relevant information, often through quantile mapping, machine learning, or hybrid approaches. Bias correction methods such as quantile mapping have been shown to improve local temperature and precipitation projections and support downstream applications~\citep{teutschbein2012bias,teng2015does,schoof2016projecting}, while statistical and dynamical downscaling address the mismatch between coarse model resolution and local-scale processes~\citep{wilby1997downscaling,jacobeit2014statistical}. However, these approaches face well-documented challenges in representing extremes, handling non-stationary biases, and preserving temporal structure, with performance varying across methods and applications~\citep{teng2015does,adachi2020methodology,yan2021updating,lazoglou2019review}.}
Alternative frameworks include quantile regression neural networks, automated regression-based statistical downscaling \citep{burger_downscaling_2012}, Gaussian mixture models \citep{ebtehaj_preservation_2010}, parameter conditioning on coarse-scale products \citep{hundecha_statistical_2008}, and variational methods for precipitation downscaling \citep{foufoula-georgiou_downscaling_2014}. 
However, these approaches lack the theoretical justification provided by extreme value theory (EVT) and are unable to extrapolate beyond the range of observed values.

{EVT has been applied in data fusion to enhance the modelling of extremes. 
For example, \citet{friederichs_statistical_2010} used ERA4 data to model the conditional 95th quantiles of precipitation in Germany, fitting a generalised Pareto distribution to threshold exceedances. 
This approach improved uncertainty estimates over non-parametric quantile regression, particularly in the upper tail. 
\citet{engelke_extremal_2019} explored the extremal behaviour of spatially aggregated data versus individual points, introducing the $\ell$-extremal coefficient to quantify spatial dependence and developing a method for statistical downscaling of temperature maxima under stationary conditions. 
However, \citet{maraun_statistical_2018} highlighted that uniform tail behaviour assumptions often fail in real-world scenarios.
Other approaches, such as those by \citet{pereira_calibration_2019} and \citet{turkman2021calibration}, use the extended generalised Pareto distribution (eGPD; \citealt{naveau_modeling_2016}) for quantile matching calibration, aligning coarser-scale data with \emph{in situ} observations. 
While the eGPD models the entire distribution and avoids threshold selection, it can introduce bias in the non-extreme component when very extreme values are present, increasing uncertainty and reducing reliability. 
Quantile matching is also sensitive to sampling variability, and scaling to larger regions requires additional assumptions about the underlying spatiotemporal structure. }

Here, we propose a data fusion model for extremes that extends the hierarchical spatiotemporal data fusion model of \citet{wilkie_nonparametric_2019}. 
The model targets threshold exceedances of AQGs while maintaining their temporal structure by censoring non-threshold exceedances and accounting for them in the likelihood.
{Specifically, we use} a zero-inflated modelling adjustment to the generalised Pareto distribution known as the Dirac-delta generalised Pareto distribution (\dGPD{}), {and perform data fusion by linking \emph{in situ} measurements and gridded CAMS reanalysis \rev{data} through the scale parameter of non-stationary \dGPD{}s.
\rev{The end product of our approach is a spatio-temporal extremal dataset over the study region, in which the tail behaviour of the gridded background field has been calibrated to reflect local extremal characteristics observed at \emph{in situ} stations.}
\rev{Our approach can be viewed as a tail-calibrated extension of statistical downscaling frameworks, in which the goal is not only to predict local values from coarse-scale inputs, but to construct a fused spatio-temporal representation with accurately calibrated extremal behaviour.}
\rev{We use our model} to describe and predict censored threshold exceedances of \PMp{} pollution in the Greater London region by using CAMS atmospheric composition reanalysis (CAMSRA/EAC4) and \emph{in situ} observation stations from the automatic urban and rural network (AURN). 
{The spatio-temporal predictive ability of our model outperforms that of the \rev{CAMSRA} data and the original method of~\cite{wilkie_nonparametric_2019}} \rev{when applied to log-transformed measurements}, and is able to capture greater variability and spatial patterns, such as higher \PMp{} concentrations near coastal areas, which are not evident in the gridded reanalysis \rev{data} alone.}

{Our key contributions lie in our model's ability to (1) preserve the temporal structure of \rev{threshold exceedance episodes while retaining information on non-exceedance occurrences through censoring}, (2) seamlessly integrate data with varying spatio-temporal resolutions while fully accounting for parameter uncertainties, and (3) \rev{generate a complete spatio-temporal dataset by calibrating the tail behaviour of gridded CAMS reanalysis data to local threshold exceedances, guided by \emph{in situ} measurements, while preserving broader spatial and temporal coverage.}}

{While the model's performance is constrained by the quality of \rev{the gridded reanalysis field} (as is the case with any data fusion approach), its flexibility allows for future enhancements, including the integration of covariates and alternative models for the full distribution. }

The remainder of this paper is structured as follows. 
{Section~\ref{sec:data} introduces the data that motivates our modelling approach.
Section~\ref{sec:Methods} outlines the proposed model and details the inference procedure.
Section~\ref{sec:application} presents a case study on air quality monitoring in the Greater London region, showcasing the results and comparing them with alternative modelling approaches.
Finally, Section~\ref{sec:DF_DandC} discusses the findings and concludes the paper.}

{The data used in the paper as well as the C++ and R codes to implement our model can be obtained from \url{https://github.com/danicuba-stats/DataFusion_for_Extremes}.}

\section{Data}\label{sec:data}
\PMp{} \emph{in situ} data in the UK are provided by the Automatic Urban and Rural Network (AURN), a nationwide network of observation stations for monitoring atmospheric composition. The AURN was established to assess compliance with the UK's Ambient Air Quality Directives (\url{https://uk-air.defra.gov.uk/networks/network-info?view=aurn}). 
Of the 171 active stations, 26 are located in the Greater London area, representing a higher spatial density compared to other regions.
For this study, we focus on data from the year 2022, the most recent publicly available dataset from the AURN. Among the 26 stations in Greater London, only 12 recorded observations for more than 75\% of the days in 2022. 

\rev{Gridded CAMS atmospheric composition reanalysis data} for \PMp{} were obtained from the ECMWF Atmospheric Composition Reanalysis 4 (EAC4) dataset, from the Copernicus Atmosphere Monitoring Service (CAMS; \url{https://ads.atmosphere.copernicus.eu/datasets/cams-global-reanalysis-eac4?tab=overview}). 
\rev{More precisely, EAC4 is a CAMS atmospheric composition reanalysis product (CAMSRA), which we use as a gridded background field. Its native horizontal resolution is approximately $0.75^\circ$, i.e., around 80 km over the study region. Alternative higher-resolution CAMS near-real-time products are available; however, CAMSRA provides a stable, fully documented baseline that is straightforward to reproduce and extend to multi-year analyses, which motivates our choice.} \rev{It is important to note that, for aerosols and near-surface PM, this product should be interpreted as a model-based reanalysis informed by data assimilation, rather than as a direct observational analogue of the \emph{in situ} measurements.}

\begin{figure}
    \centering
    \includegraphics[scale=0.45]{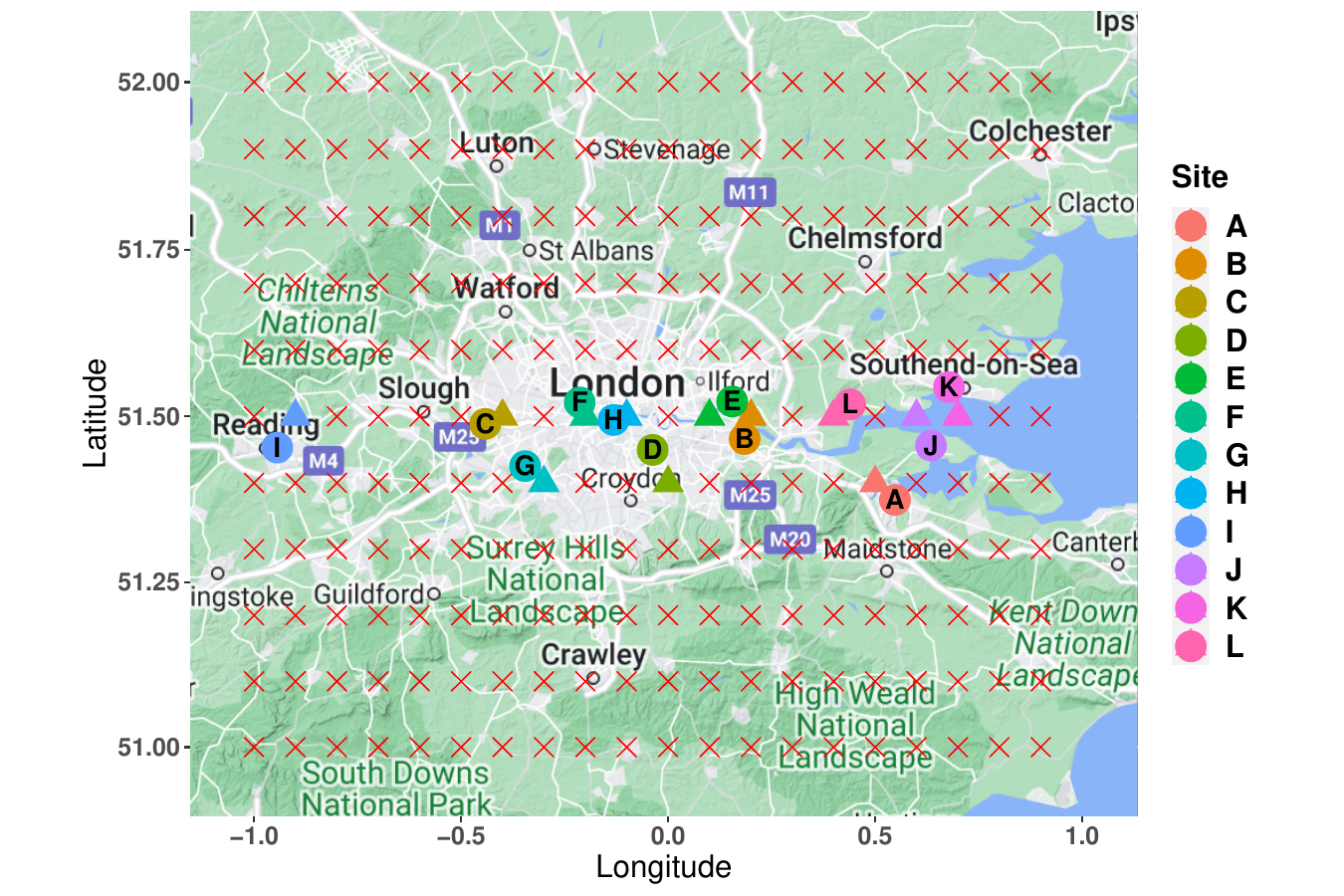}
    \caption{Coloured circles denote the locations of the AURN observation stations, while coloured triangles indicate the nearest centroid in the CAMSRA grid. The red crosses indicate the remaining CAMSRA grid centroids. }
    \label{fig:DF_AURN_EAC4_map}
\end{figure}

{To align with air quality guidelines (AQGs), data from both sources were aggregated into 24-hour means. 
Figure~\ref{fig:DF_AURN_EAC4_map} illustrates the locations of both data sources: crosses represent centroids of the CAMSRA grid, coloured circles mark the locations of the AURN observation stations, and coloured triangles indicate the nearest CAMSRA grid centroid to each of the 12 AURN stations.}

\begin{figure}
    \centering
        \includegraphics[scale=0.5]{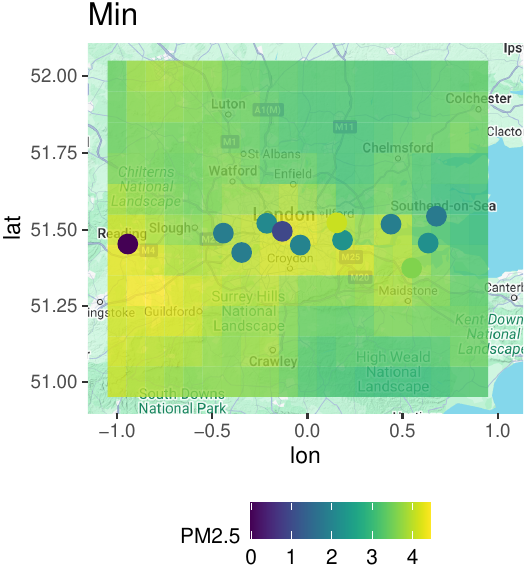}
        \includegraphics[scale=0.5]{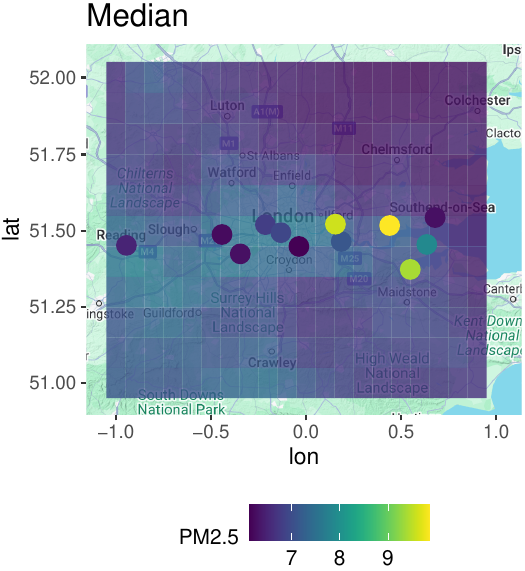}
        \includegraphics[scale=0.5]{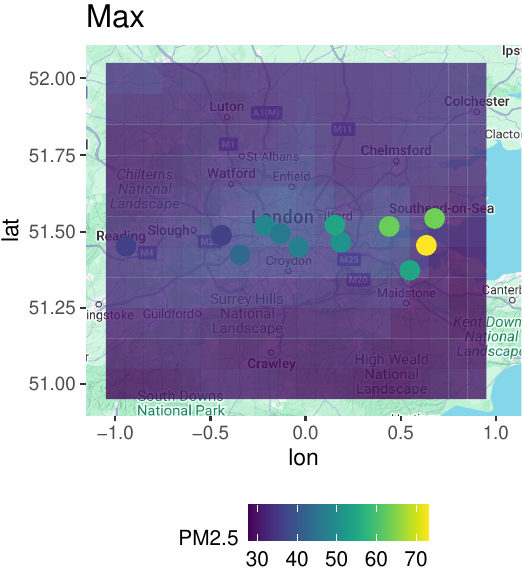}
    \caption{Maps of the minimum, median and maximum of the CAMSRA data and AURN observation stations. }
    \label{fig:DF_all_spat}
\end{figure}

\begin{figure}
    \centering
    \includegraphics[scale=0.65]{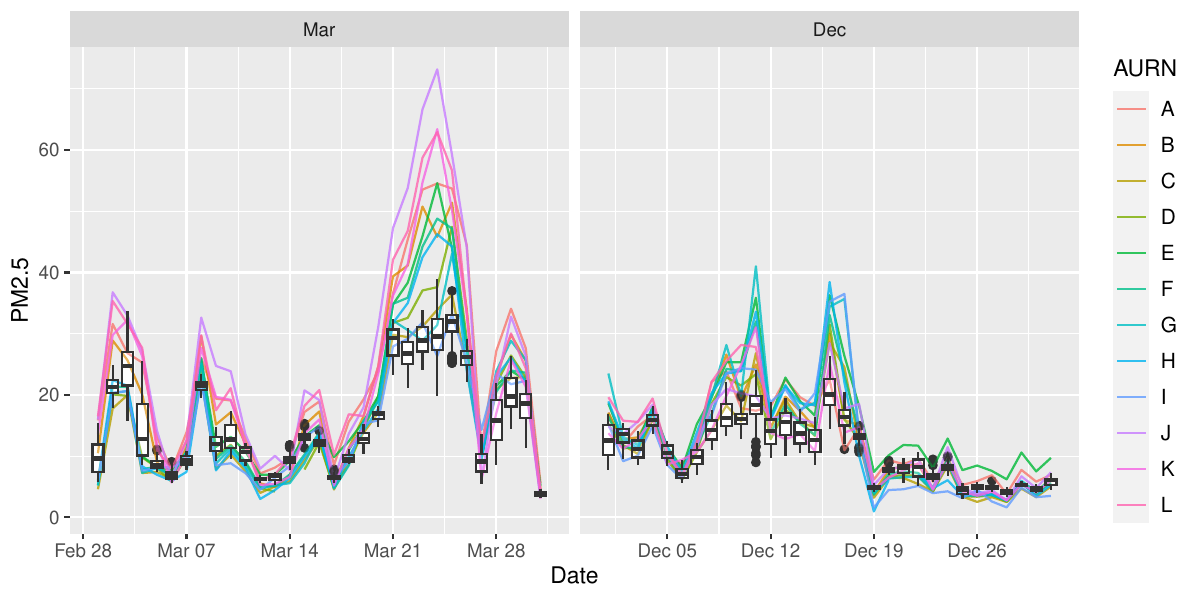}
    \caption{Temporal trends of the AURN and CAMSRA data for March and December. Lines indicate daily concentrations of the 12 AURN sites, following the colour pattern in Figure~\ref{fig:DF_AURN_EAC4_map}. Boxplots represent summarised CAMSRA data at the coloured triangles in Figure~\ref{fig:DF_AURN_EAC4_map} (i.e., nearest centroids to the AURN locations).}
    \label{fig:NS_DF_CAMS_marchdec}
\end{figure}

While the CAMSRA data have complete spatial and temporal coverage of the Greater London region, the observations are known to be overly smooth, often displaying significant discrepancies compared to \emph{in situ} measurements even after calibration. {These discrepancies are known to increase }with extreme values, particularly in regions far from the equator and near the coast, where the largest biases are observed \citep{sheridan_comparison_2020}. 
{Figure \ref{fig:DF_all_spat} shows the location-wise minimum, median and maximum for both datasets.
The differences between the data sources are most pronounced for the minimum and maximum values: CAMSRA overestimates minimum and underestimates maximum values, with the largest discrepancies occurring at stations near the east coast. }
{A similar pattern is observed over time. Figure \ref{fig:NS_DF_CAMS_marchdec} shows temporal trends of the AURN and CAMSRA data for March and December, months that exhibit the highest peaks in the AURN dataset. In the figure, daily concentrations of the 12 AURN sites are given as lines (following the same colour palette used in Figure~\ref{fig:DF_AURN_EAC4_map}), while CAMSRA data associated with the nearest centroids to the AURN locations are summarised using daily boxplots.
We can see that the AURN peaks are generally underestimated by the CAMSRA data, with the discrepancy being most pronounced at sites located further from central London, such as locations A, B, G, and I to L, as shown in Figure~\ref{fig:DF_all_spat}.
Similar plots for all months are provided in Figure~\ref{fig:appendix_NS_DF_CAMS_ts1} of the Supplementary Material.}

\begin{figure}
    \centering
    \includegraphics[scale=0.35]{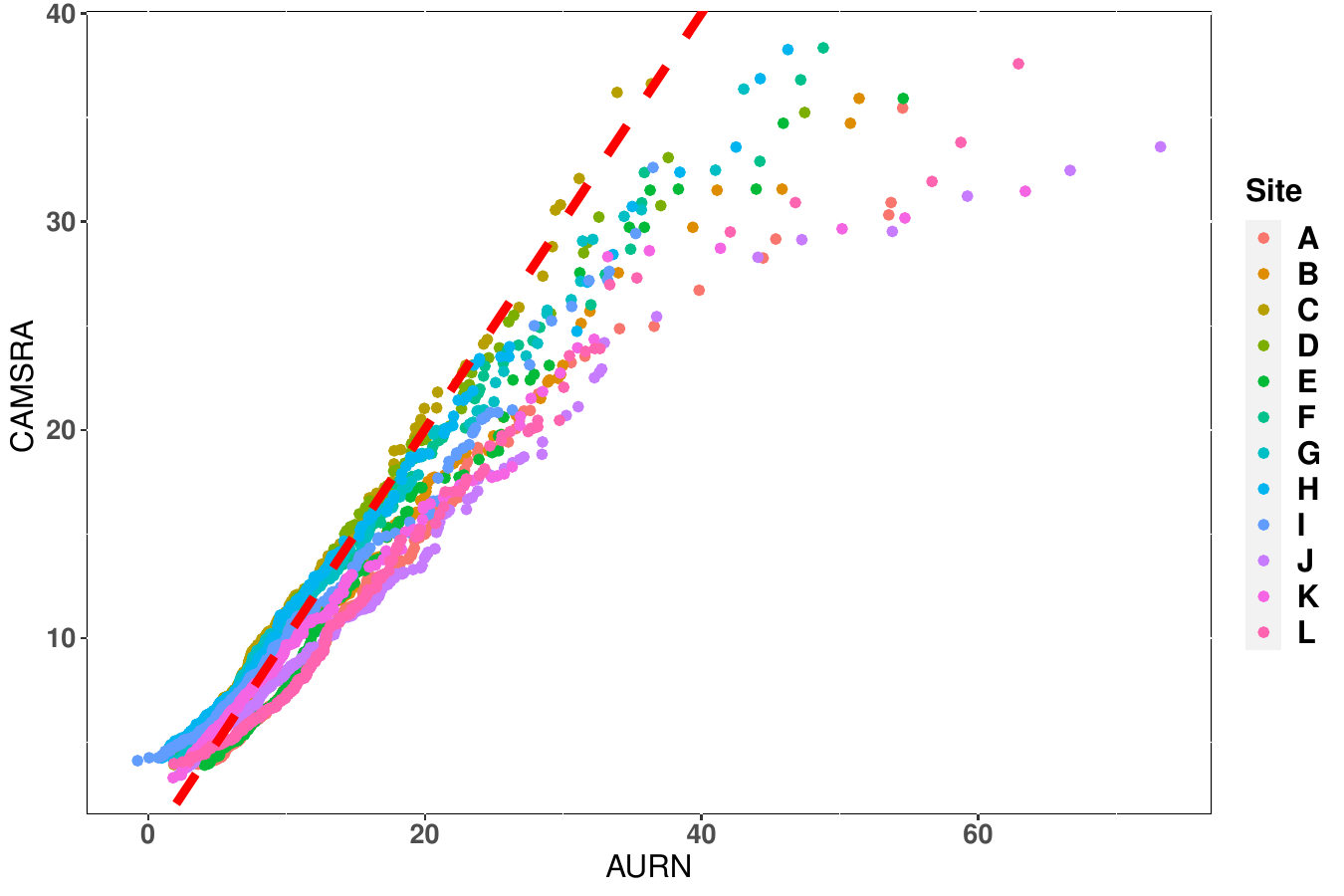}
    \caption{Q-Q plot comparing data from each AURN site with the corresponding nearest grid centroid from the CAMSRA dataset. }
    \label{fig:DF_AURN_EAC4_QQ}
\end{figure}

{Figure~\ref{fig:DF_AURN_EAC4_QQ} compares quantiles of the AURN data to those of the CAMSRA data at the nearest grid centroids to the AURN locations (triangles in Figure~\ref{fig:DF_AURN_EAC4_map}). As expected, the two data sources are highly correlated, but two key discrepancies are apparent.
First, in the bottom-left corner of the plot, most sites lie above the 45$^\circ$ reference line, indicating that CAMSRA overestimates very small values, consistent with observations from Figures~\ref{fig:DF_all_spat} and~\ref{fig:NS_DF_CAMS_marchdec}. Second, for large values, the CAMSRA data consistently underestimates AURN measurements, particularly at sites further from central London, such as A, B, G, and I to L.
Notably, site C shows closer alignment with the 45$^\circ$ line, reflecting better agreement between the CAMSRA and AURN datasets for that location.}

{Our study focuses on threshold exceedances, so an appropriate threshold must be defined prior to fitting the model that we propose in Section~\ref{sec:Methods}. The tradeoff between bias and variance in threshold selection for GPD-based models is well-documented (see, e.g., \citealp{scarrott2012review}), with most threshold selection tools being graphical in nature. However, recent efforts have aimed at developing more automated methods that account for threshold estimation uncertainty~\citep{murphy2024automated}.
Here, we use the stability of mean residual life plots~\citep[Ch.~4]{coles_introduction_2001} and find that the 80\% station-wise empirical quantile of the AURN data provides a reasonable fit, retaining approximately 75 observations per station. 
For consistency, we use the same quantile for the CAMSRA data.}

\section{Methodology}\label{sec:Methods}
\subsection{Modelling framework}\label{sec:modelframework}
Spatial and spatio-temporal data fusion models are common in the literature and span a wide range of approaches. 
\citet{wilkie_data_2015} proposed a Gaussian downscaling data fusion model for monitoring water quality in Lake Balaton, Hungary, based on the spatially-varying parameter approach of \citet{gelfand_spatial_2003} and \citet{berrocal_spatio-temporal_2010}. 
Under this framework, the true underlying process $Y$ is not assumed but rather sampled by the \emph{in situ} measurements at a finite number of sample locations, $\boldsymbol{s} \in \mathcal{S} \subset \mathbb{R}^2$, which are linked to the \rev{gridded background data} at the same locations, $x(\boldsymbol{s})$, via a linear regression model with spatially varying coefficients.
The model is defined as 
\begin{equation*}
    Y(\mathbf{s}) = \mu(\mathbf{s}) + W(\mathbf{s}) + \epsilon(\mathbf{s}),
\end{equation*}
where $\mu(\mathbf{s})$ is a function of the \rev{gridded data} {$x(\mathbf{s})$} at the same location as the \emph{in situ} sensors (i.e., \emph{collocated} data).
It is expressed as $\mu(\mathbf{s}) = \alpha(\mathbf{s}) + \beta(\mathbf{s})x(\mathbf{s})$, where $\alpha(\mathbf{s})$ and $\beta(\mathbf{s})$ are spatially-varying parameters.
$\epsilon(\mathbf{s})$ represent independent, normally distributed error terms with mean $0$ and variance $\tau^2$, and $W(\mathbf{s})$ is a second-order stationary process centred at zero that is independent of $\epsilon(\mathbf{s})$. 
\rev{In this context, data fusion links gridded background information to the mean of the true process $Y$.}
This relationship allows for the prediction of $Y$ at a location where no observation station is present. 
\citet{wilkie_nonparametric_2019} extended the model to a spatio-temporal framework, formulating it as a hierarchical Bayesian approach. 
In this extension, the coefficients vary across both time and space, with temporal trends captured using basis functions and spatial dependence described through an exponential decay term.

We aim to adapt the approach of \citet{wilkie_nonparametric_2019} to account for extremes, which requires a precise definition of extreme values. 
Following the established practice in AQG-based studies, we define extremes as exceedances over a high threshold and draw on a key result from univariate extreme value theory.
This result characterises the distribution of threshold excesses using the generalised Pareto distribution (GPD).
Specifically, for a random variable $Z$ and a sufficiently large threshold $u$, the distribution of the threshold exceedances, $(Z-u)$, conditional on $(Z>u)$, is asymptotically approximated by the GPD with distribution function given by
\begin{equation}\label{eq:SB_GPD}
    H(z) = 1-\left(1+\frac{\xi z}{{\sigma_u}}\right)^{-1/\xi},
\end{equation}
defined on $\{z:z>0 \text{ and } (1+\xi z/{\sigma_u})>0\}$.
In~\eqref{eq:SB_GPD}, $\sigma_u>0$ is the threshold-dependent scale parameter and $\xi\in\mathbb{R}$ is the shape parameter.
{The approximation in~\eqref{eq:SB_GPD} is valid under max-stability assumptions; see, e.g., \citealp[Ch.~4]{coles_introduction_2001} for details.}
The GPD gained significant popularity since the publication of {\citet{davison_models_1990}} where it was used in the context of generalised linear models.
Since then, there have been multiple applications of the GPD in a variety of contexts, from financial {\citep{gilli_application_2006,furio_extreme_2013,bufalo_addressing_2024}}, to environmental {\citep{huang2019estimating, sharkey2019bayesian, castro2021bayesian}} and engineering {\citep{youngman_geostatistical_2016,ross_efficient_2017, mackay_assessment_2020}} applications.

The GPD allows us to characterise threshold excesses at each location. 
So, at a given location and once a suitable high threshold $u$ is defined, fitting the GPD involves discarding observations that do not exceed the threshold, producing temporal gaps.
The use of the GPD for spatio-temporal predictions is, therefore, limited since it produces a time series that is not continuous in time.
A naive extension of \citet{wilkie_nonparametric_2019} approach would involve replacing the Gaussian likelihood with that of the GPD, but this would result in an unrealistic time series exclusively comprised of threshold exceedances.
To address this issue, we could directly model non-threshold exceedances, resulting in a data fusion model for both bulk and tail values.
Alternatively, we could account for the presence of non-exceedances via a censoring scheme. 
Here, we adopt the latter approach.
{To describe our censoring method, let $\boldsymbol{y}_i = (y_{ij_0},\ldots,y_{ij_i})^\top$ be the \emph{in situ} data at location $i=1,\ldots,n$, and let $\boldsymbol{x}_i = (x_{ik_0},\ldots,x_{ik_i})^\top$ denote the \rev{gridded CAMSRA data} for the grid cell {with the nearest centroid to the} \emph{in situ} location $i$. 
\rev{This nearest-centroid matching follows a pragmatic collocation principle that is common in this literature. For example, in the air-quality downscaling framework of \citet{berrocal_spatio-temporal_2010}, each monitoring point is associated with the grid cell in which it lies, and \citet{berrocal2010bivariate} adopt the same strategy in their bivariate space-time downscaler under spatial misalignment. Our use of a single matched grid location provides a simple and interpretable way to connect point-referenced observations with a gridded background field while keeping the model specification and computation manageable.}}
\rev{More elaborate collocation schemes, such as using several surrounding grid cells or interpolation-based approaches, could certainly be considered. For instance, \citet{riley2025bayesian} reproject coarse gridded variables onto a 1 km grid using inverse distance weighting before fitting their model on a common support.}
Note that we allow the \rev{gridded reanalysis data} to have different time points than the \emph{in situ} data. In most cases, \rev{these measurements} either begin at the same time as the \emph{in situ} measurements ($k_0 = j_0$) or earlier ($k_0 < j_0$). Similarly, \rev{gridded reanalysis data} often extend over the same period as the \emph{in situ} data ($k_i = j_i$), although in some cases they might cover a shorter or longer time span ($k_i < j_i$ or $k_i > j_i$, respectively).
The censored observations of $\boldsymbol{y}_i$ and $\boldsymbol{x}_i$ are denoted by $\boldsymbol{y}^*_i = (y^*_{ij_0},\ldots,y^*_{ij_i})^\top$ and $\boldsymbol{x}^*_i = (x^*_{ik_0},\ldots,x^*_{ik_i})^\top$, respectively, and are defined as
\begin{equation}\label{eq:DF_Nonthreshold}
    \begin{aligned}
    {y}^{*}_{ij} & =
    \begin{cases}
        0, & \quad \text{if } y_{ij} \leq u_{y_i},\\
        y_{ij} - u_{y_i}, & \quad \text{if } y_{ij} > u_{y_i},\quad j = j_0,\ldots, j_i,\\
    \end{cases}\\
    {x}^{*}_{ik} & =  
    \begin{cases}
        0, & \quad \text{if } x_{ik} \leq u_{x_i},\\
        x_{ik} - u_{x_i}, & \quad \text{if } x_{ik} > u_{x_i},\quad k = k_0,\ldots, k_i,\\
    \end{cases}
    \end{aligned}
\end{equation}
where $u_{y_i}$ and $u_{x_i}$ represent high enough site-specific threshold, such that
the GPD in~\eqref{eq:SB_GPD} is a suitable approximation of the behaviour of threshold exceedances $y_{ij} - u_{y_i}\mid y_{ij} > u_{y_i}$ and $x_{ik} - u_{x_i}\mid x_{ik} > u_{x_i}$. 
Following the approaches by \citet{weglarczyk_three-parameter_2005} and \citet{couturier_zero-inflated_2010}, we describe 
$\boldsymbol{y}^*_i$ and $\boldsymbol{x}^*_i$ using 
a zero-inflated GPD mixture model called the Dirac-delta generalised Pareto distribution (\dGPD{}).
The density of the \dGPD{} is defined as 
\begin{equation} \label{eq:DF_dGPD}
    f(z|p,\sigma,\xi) = (1-p_u)\delta(z) + \frac{p_u}{\sigma_u} \left(1 + \frac{\xi z}{\sigma_u}\right)^{-1/\xi -1}\Delta_0(z),
\end{equation}
where $\sigma_u$ and $\xi$ are, respectively, the scale and shape parameters of the GPD in~\eqref{eq:SB_GPD}, $p_u \in [0,1]$ is the probability of exceedances over the threshold $u$, $\delta(z)$ is the Dirac delta function with point mass only at $z=0$, and $\Delta_0(z)$ is the unit step function, equalling $1$ when $z>0$. 
Using this likelihood, we can account for non-threshold exceedances, which contribute $1-p_u$ to the likelihood, effectively allowing us to predict the occurrence of non-extremes.

In the spirit of \citet{couturier_zero-inflated_2010}, we extend the model in \eqref{eq:DF_dGPD} to include covariates in the exceedance probability of \emph{in situ} measurements through a GLM framework. 
\rev{Specifically, we use lagged indicators of exceedances in the gridded data as covariates to model exceedance probabilities in the \emph{in situ} series.}
{For time $j$ at location $i$, we denote the exceedance probability of \emph{in situ} measurements as $p_{u,y_{ij}}$ and assume the logistic model
\begin{equation}\label{eq:DF_logitreg}
    \text{logit}\left(p_{u,y_{ij}}\right) = \mathbf{W}_i[j,\cdot]\boldsymbol{\lambda}_{i},\quad j=j_0,\ldots,j_i,
\end{equation}
where $\boldsymbol{\lambda}_{i}$ is a location-specific vector of coefficients, $\mathbf{W}_i$ is a covariate matrix for location $i$, which encodes the information about threshold exceedances in the corresponding \rev{gridded CAMSRA data}, $\boldsymbol{x}_i^*$. 
Specifically, each row of $\mathbf{W}_i$ corresponds to a time point $j = j_0, \ldots, j_i$, and the columns represent covariates based on the threshold exceedances for times $j-1$, $j$, and $j+1$ (when available). 
The matrix is structured as follows: the first column is a constant vector, the second column corresponds to exceedances at time $j-1$, the third column corresponds to exceedances at time $j$, and the fourth column corresponds to exceedances at time $j+1$. 
The covariate matrix $\mathbf{W}_i$ can be expressed as:
\begin{equation*}
\mathbf{W}_i=\left[\begin{array}{cccc}
\mathbf{1} & \mathbf{w}_{i}^{(1)} & \mathbf{w}_{i}^{(2)} & \mathbf{w}_{i}^{(3)} \\
\end{array}\right],
\end{equation*}
where $\mathbf{1}$ is a column vector of ones, and $\mathbf{w}_i^{(l)} = (w_{ij_0}^{(l)},\ldots,w_{ij_i}^{(l)})^\top$, $l=1,2,3$ are the indicator vectors for exceedances at times $j-1$, $j$, and $j+1$, respectively.
Specifically, for $j=j_0,\ldots,j_i$, we have
\begin{equation}\label{eq:Wcomponents}
    \begin{aligned}
    w_{ij}^{(1)} & =
    \begin{cases}
        \mathbbm{1}(x_{i(j-1)}^* >0) & \quad \text{ if } j-1\geq k_0,\\
        0, & \quad \text{ otherwise},\\
    \end{cases}\\
     w_{ij}^{(2)} & = \mathbbm{1}(x_{ij}^* >0), \\
    w_{ij}^{(3)} & =
    \begin{cases}
        \mathbbm{1}(x_{i(j+1)}^* >0) & \quad \text{ if } j+1\leq k_i,\\
        0, & \quad \text{ otherwise}.\\
    \end{cases}
    \end{aligned}
\end{equation}}
The interpretation of the matrix $\mathbf{W}_i$ is that each row corresponds to a specific time point, and the covariates are the indicators of whether there was a threshold exceedance in the \rev{reanalysis data} at the corresponding time points.
{So, for instance, if the \rev{gridded reanalysis} and \emph{in situ} measurements begin at the same time ($k_0=j_0=1$) and cover the same time span up to some time point $J$ ($k_i=j_i=J$), the matrix $\mathbf{W}_i$ would be given by
\begin{equation*}
\mathbf{W}_i=\left[\begin{array}{cccc}
1 & 0 & \mathbbm{1}({x}^{*}_{i1}>0) & \mathbbm{1}({x}^{*}_{i2}>0) \\
1 & \mathbbm{1}({x}^{*}_{i1}>0) & \mathbbm{1}({x}^{*}_{i2}>0) & \mathbbm{1}({x}^{*}_{i3}>0) \\
. & \cdot & . & \cdot \\
\cdot & \cdot & \cdot & \cdot \\
\cdot & \cdot & \cdot & \mathbbm{1}({x}^{*}_{iJ}>0) \\
1 & \mathbbm{1}({x}^{*}_{i(J - 1)}>0) & \mathbbm{1}({x}^{*}_{iJ}>0) & 0
\end{array}\right],
\end{equation*}
and the vector of coefficients would be {$\boldsymbol{\lambda}_{i} = (\lambda_{i0},\lambda_{i1},\lambda_{i2},\lambda_{i3})^\top$.}}
{We can see that the matrix $\mathbf{W}_i$ at time point $j$ (specifically, the row $\mathbf{W}_i[j,\cdot]$) incorporates indicators of threshold exceedances in $\boldsymbol{x}_i$ at times $j-1$, $j$, and $j+1$ (when available) as covariates to predict $\boldsymbol{y}_i^*$ at time $j$ (i.e., to predict whether $y_{ij}$ is a threshold exceedance).}
This choice of $\mathbf{W}_i$ has a physical interpretation, as it is not uncommon for \rev{gridded reanalysis or other model-based background data} to have a lagged reaction to physical occurrences.

Exceedance probabilities {$p_{u,x_{ik}}$, $k=k_0,\ldots,k_i$, associated with the \rev{gridded CAMSRA measurements} $\boldsymbol{x}_i^*$ are characterised by their empirical probabilities. 
This simpler approach provides a way to estimate these probabilities, leveraging the completeness of records without increasing model complexity. 
\rev{To avoid ambiguity, we stress that the gridded $x$-series is treated as observed background input rather than as a second latent process to be modelled recursively. Accordingly, $p_{u,x_{ik}}$
is the empirical exceedance proportion computed from the thresholded gridded series, and is used as a plug-in quantity in the $\delta$-GPD specification.}

{As we mentioned above, we aim to model $\boldsymbol{y}^*_i$ and $\boldsymbol{x}^*_i$ using the \dGPD{} defined in~\eqref{eq:DF_dGPD}, which requires specifying our approach to estimate the shape and scale parameters.
\rev{We assume constant shape parameters $\xi_y\in\mathbb{R}$ and $\xi_x\in\mathbb{R}$ across space and time, supported by exploratory \emph{in situ} fits showing no substantial differences at most locations, and motivated by identifiability and parsimony considerations given the flexible spatio-temporal structure already included in the scale component.}
\rev{This choice also} aligns with the methodology of \cite{wilkie_nonparametric_2019}, where the Gaussian variance is treated as a constant, independent hyperparameter for each data source.}
Additional constraints can be imposed on $\xi_y$ and $\xi_x$ to ensure desirable properties of the \dGPD{}. In this paper, we focus on preserving the first and second moments of the \dGPD{} as well as ensuring the regularity of maximum likelihood estimators (MLEs). 
We found that a prior on $\xi_y$ and $\xi_x$ defined as a scaled Laplace distribution on $(-0.5,0.5)$ has the desired effect while also promoting model parsimony by encouraging simpler models.
See Section~\ref{sec:priors_data} for more details and justification of our prior selection.

{The \dGPD{} scale parameters warrant special attention, as they encapsulate spatio-temporal information and play a key role in our data fusion approach.
We model them as smooth temporal curves, fitted using basis functions of the same dimension. \rev{The resulting non-stationarity is therefore smooth rather than abrupt: daily variation is represented by evaluating low-rank temporal basis expansions at each day, with the site-specific coefficients determining the local shape of those curves.}
Specifically, we assume that
\begin{equation}\label{eq:scales}
    \log\{\sigma_{u_y,i}\} = \Phi_i \boldsymbol{c}_i,\qquad \log\{\sigma_{u_x,i}\} = \Psi_i \boldsymbol{d}_i, \quad i = 1,\ldots,n,
\end{equation}
where $\Phi_i$ and $\Psi_i$ are matrices of basis functions of dimensions $j_i \times m$ and $k_i \times m$, respectively.
\rev{In the implementation used in Section~\ref{sec:application}, temporal variation is represented using a cubic B-spline basis (order 4) with $m$ basis functions, constructed over the observed time range with equally spaced knots.}}
\rev{Their purpose is to provide a low-dimensional smooth representation of temporal variation in the \dGPD{} scale parameters, rather than to estimate a separate parameter at each day. Thus, the temporal pattern is not fixed in a preliminary smoothing step, but is learned jointly from the data through the posterior distribution of the basis coefficients. To borrow strength across sites, the location-specific basis coefficients are assigned a distance-based spatial dependence structure, so that nearby sites are encouraged to have more similar temporal evolution while still allowing local differences.}
The vectors of basis coefficients $\textbf{c}_i = (c_{i1},\dots,c_{im})^\top$ and $\textbf{d}_i = (d_{i1},\dots,d_{im})^\top$ are then linked linearly using spatially-varying regression coefficients.
Specifically, for $r = 1,\ldots,m$,
\begin{align}\label{eq:cijdij}
    c_{ir} = \alpha_{ir} + \beta_{ir}d_{ir} + \epsilon_c&,\qquad \epsilon_c\mid \sigma_c^2\sim \mathrm{N}(0,\sigma_c^2)\nonumber\\
    \boldsymbol{\alpha}_r = (\alpha_{1r},\ldots,\alpha_{nr})^\top
 \mid \sigma_\alpha^2 & \sim \mathrm{N}_n\left(\boldsymbol{0}, \sigma_\alpha^2 \exp \left(-\phi_\alpha \Sigma_{\mathrm{data}}\right)\right),\nonumber \\
 \boldsymbol{\beta}_r = (\beta_{1r},\ldots,\beta_{nr})^\top \mid \sigma_\beta^2 & \sim \mathrm{N}_n\left(\boldsymbol{1}, \sigma_\beta^2 \exp \left(-\phi_\beta \Sigma_{\mathrm{data}}\right)\right),
\end{align}}
{where $\Sigma_{\mathrm{data}}$ is the $n\times n$ matrix of Euclidean distances between the $n$ point locations of the \emph{in situ} data and $\sigma_\alpha^2$ and $\sigma_\beta^2$ are the spatial variances.
The parameters $\phi_\alpha$ and $\phi_\beta$ are spatial decay parameters that control the rate at which the correlations of the intercept and slope parameters diminish as the distance between locations increases.}
\rev{This approach effectively imposes a non-stationary model for the threshold sizes. Here, non-stationarity at the daily scale does not mean that each day is assigned an independent extreme-value law. Rather, the temporal basis expansion imposes smooth variation, so that nearby days borrow strength from one another and the exceedance behaviour evolves gradually in response to slowly varying background conditions, such as seasonal progression and persistent meteorological changes.}

{The model components in~\eqref{eq:scales} and~\eqref{eq:cijdij} are central to our spatio-temporal data fusion framework. 
Indeed, the basis matrices $\Phi_i$ and $\Psi_i$ do not necessarily have the same number of rows, therefore enabling flexibility in temporal support.
This ensures our model can be fitted in common scenarios where \emph{in situ} observations are unavailable at certain time points while \rev{gridded reanalysis data} remain available.
The two data sources are seamlessly integrated through a Gaussian linear model that links the basis coefficient vectors $\textbf{c}_i$ and $\textbf{d}_i$. 
Additionally, spatially varying regression coefficients capture and encode spatial correlations among \emph{in situ} locations.
In essence, the model components in~\eqref{eq:scales} and~\eqref{eq:cijdij} are the cornerstone of our approach, enabling the fusion of data with varying spatio-temporal resolutions.}

\rev{Finally, it is important to note that, under our specification, spatial dependence is captured primarily through the latent spatio-temporal structure assigned to the \dGPD{} scale parameter. The exceedance-probability component does not include an additional spatial random effect, and therefore any residual spatial clustering in the occurrence process is not modelled explicitly. We adopt this simplification to preserve identifiability and computational tractability in the present small-network setting.}
\rev{Additionally, although the two data sources do not play symmetric roles, the aim of the model is not simply to predict large \emph{in situ} values given gridded input. Rather, it is to construct a fused spatio-temporal extremal representation whose tail behaviour is calibrated by combining the local accuracy of the \emph{in situ} observations with the broader spatio-temporal coverage of the gridded background field.}

{We now present our complete spatio-temporal extreme data fusion approach (ExDF) within a hierarchical framework. 
The model in~\eqref{eq:DF3} is written in a self-contained manner, so understanding it fully requires referencing previously introduced components, which may involve some repetition of earlier details.
{Recall that $\boldsymbol{y}^*_i = (y_{ij_0}^*,\ldots,y_{ij_i}^*)^\top$ represents the temporal vector of \emph{in situ} observations at location $i$, while $\boldsymbol{x}^*_i = (x_{ik_0}^*,\ldots,x_{ik_i}^*)^\top$ corresponds to the \rev{gridded CAMSRA data} for the grid cell with the centroid nearest to the \emph{in situ} location $i$}. 
The ExDF model is structured as follows:}
\begin{equation}\label{eq:DF3}
    \begin{aligned} 
    \boldsymbol{y}^{*}_i\mid \boldsymbol{c}_i & \sim \delta\text{-GPD}\left(\text{exp}(\Phi_i \boldsymbol{c}_i), \xi_y, \mathbf{p}_{yi}\right),\\
    \text{logit}\left(p_{u,y_{ij}}\right) &= \mathbf{W_i}[j,\cdot]\boldsymbol{\lambda}_{i}\\
    c_{i r} \mid \alpha_{i r}, \beta_{i r}, d_{i r}, \sigma_c^2 & \sim \mathrm{N}\left(\alpha_{i r}+\beta_{i r} d_{i r}, \sigma_c^2\right),\quad r = 1,\ldots,m,\\ 
    \boldsymbol{x}^{*}_i \mid \boldsymbol{d}_i & \sim \delta\text{-GPD}\left(\text{exp}(\Psi_i \boldsymbol{d}_i), \xi_x, \hat{\mathbf{p}}_{xi} \right), \\ 
    \hat p_{u,x_{ik}} &= \mathbbm{1}(x^{*}_{ik}>0),\\
    \boldsymbol{d}_i & \sim \mathrm{N}_m\left(\boldsymbol{\mu}_d, \Sigma_d\right),\\
    \xi_y & \sim \text{Laplace}_{\xi}(\mu_y,b_y) \quad \text{where } -0.5 < \xi_y < 0.5,\\
    \xi_x & \sim \text{Laplace}_{\xi}(\mu_x,b_x), \quad \text{where } -0.5 < \xi_x < 0.5,\\
    \boldsymbol{\alpha}_j = (\alpha_{1r},\ldots,\alpha_{nr})^\top \mid \sigma_\alpha^2 & \sim \mathrm{N}_n\left(\boldsymbol{0}, \sigma_\alpha^2 \exp \left(-\phi_\alpha \Sigma_{\mathrm{data}}\right)\right), \\
    \boldsymbol{\beta}_r = (\beta_{1r},\ldots,\beta_{nr})^\top \mid \sigma_\beta^2 & \sim \mathrm{N}_n\left(\boldsymbol{1}, \sigma_\beta^2 \exp \left(-\phi_\beta \Sigma_{\mathrm{data}}\right)\right), \quad r = 1,\ldots, m,\\ 
    \left(\sigma_\alpha^2\right)^{-1} & \sim \mathrm{Ga}\left(a_\alpha,b_\alpha\right),\\
    \left(\sigma_\beta^2\right)^{-1} & \sim \mathrm{Ga}\left(a_\beta, b_\beta\right), \\ 
    \left(\sigma_c^2\right)^{-1} & \sim \mathrm{Ga}\left(a_c, b_c\right),\\ 
    \lambda_{il} & \sim N(\mu_{\lambda_l},\sigma^2_{\lambda_l}),\quad l=0,1,2,3,\\
    \end{aligned}
\end{equation}
where:
\begin{itemize}[leftmargin=*]
    \item $\boldsymbol{\Phi}$ is a $j_i \times m$ matrix of basis functions evaluated at every time point of $\boldsymbol{y}_i^*$ and $\boldsymbol{c}_i = (c_{i1},\ldots,c_{im})^\top$ is the corresponding vector of basis coefficients.
    \item $\xi_y$ is the GPD shape parameter for $\boldsymbol{y}_i^*$, constant over time.
    \item $\mathbf{p}_{yi} = (p_{u,y_{ij_0}},\ldots,p_{u,y_{ij_i}})^\top$ is the vector of exceedance probabilities for every element in $\boldsymbol{y}_i^*$.
    \item $\mathbf{W_i}[j,\cdot] = (1,w_{ij}^{(1)},w_{ij}^{(2)},w_{ij}^{(3)})^\top$, with every $w_{ij}^{(l)}$, $l=1,2,3$, described as in~\eqref{eq:Wcomponents}, and {$\boldsymbol{\lambda}_{i} = (\lambda_{i0}, \lambda_{i1}, \lambda_{i2}, \lambda_{i3})^\top$} is a {location-specific vector of linear coefficients}. 
    \item $\boldsymbol{\Psi}$ is a $k_i \times m$ matrix of basis functions evaluated at every time point of $\boldsymbol{x}_i$ and $\boldsymbol{d}_i = (d_{i1},\ldots,d_{im})^\top$ is the corresponding vector of basis coefficients.
    \item $\xi_x$ is the GPD shape parameter for $\boldsymbol{x}_i^*$, constant over time.
    \item $\hat{\mathbf{p}}_{xi} = (\hat p_{u,x_{ik_0}},\ldots,\hat p_{u,x_{ik_i}})^\top$ is the vector of \rev{empirical} exceedance probabilities for every element in $\boldsymbol{x}_i^*$.
    \item $\Sigma_{\text{data}}$ is an $n \times n$ matrix of distances between the $n$ \emph{in situ} locations.
    \item $\phi_\alpha$ and $\phi_\beta$ are the spatial decay parameters. 
    \item $\boldsymbol{\alpha}_j$ and $\boldsymbol{\beta}_k$ are temporal additive and multiplicative bias between the vector of basis coefficients $\boldsymbol{c}_i$ and $\boldsymbol{d}_i$.
    \item $a_\alpha, b_\alpha, a_\beta, b_\beta, a_c$ and $b_c$ are hyperparameters related to different scale parameters, while
    $\mu_y, b_y, \mu_x,b_x, \mu_{\lambda_0}, \mu_{\lambda_1}, \mu_{\lambda_2}, \mu_{\lambda_3}, \sigma_{\lambda_0}^2,\sigma_{\lambda_1}^2,\sigma_{\lambda_2}^2$ and $\sigma_{\lambda_3}^2 $ are hyperparameters for the estimation of the shape and exceedance probability parameters. The choice of hyperparameters is application-specific and discussed further in Section \ref{sec:application}. 

\end{itemize}

\subsection{Inference}
\begin{figure}
    \centering
    \includegraphics[scale=0.75]{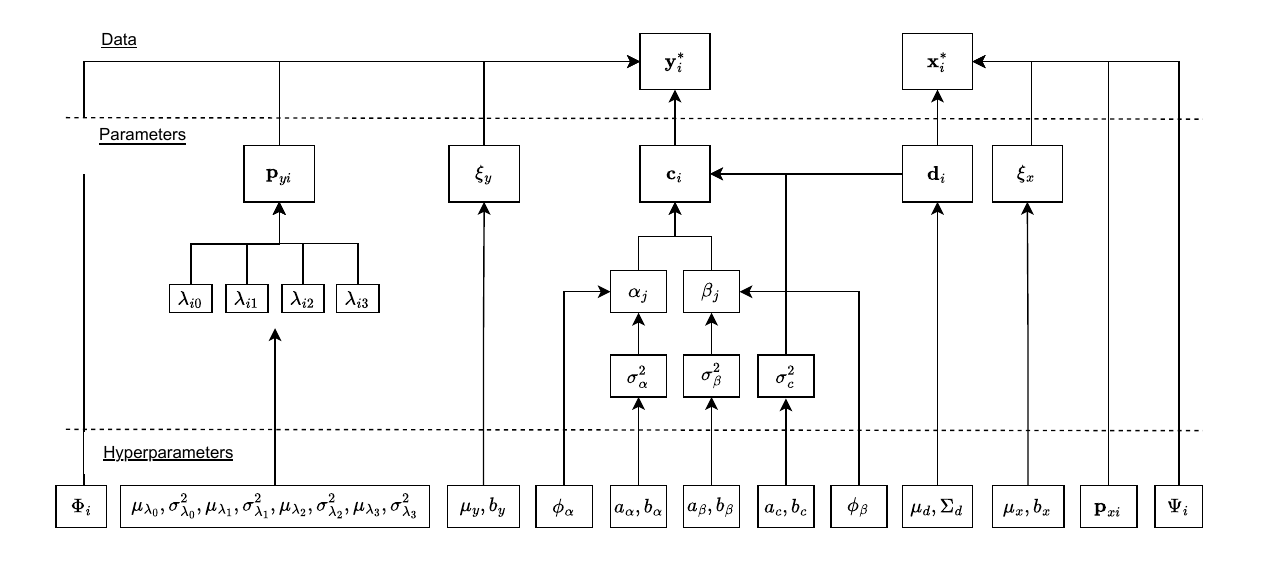}
    \caption[Flowchart of data fusion for extremes model.]{Hierarchy of the data fusion model for threshold exceedances defined in \eqref{eq:DF3} and referred to as the ExDF model.}
    \label{fig:DF_FlowChart}
\end{figure}

The model in~\eqref{eq:DF3} is fitted using Metropolis-Hastings Markov Chain Monte Carlo. 
Unlike the Gaussian approach of \citet{wilkie_nonparametric_2019}, where conjugate priors and likelihoods allow for closed-form posteriors sampled via Gibbs' sampler, the \dGPD{} likelihood does not lead to conjugate posteriors. 
Therefore, our approach requires the use of the Metropolis-Hastings sampler, where posterior samples are obtained through an acceptance/rejection process. 
{The inference procedure is illustrated in Figure~\ref{fig:DF_FlowChart}, which helps clarify the relationships between parameters and hyperparameters. 
The model was implemented in \texttt{C++} for computational efficiency and was integrated into \texttt{R} using the \texttt{Rcpp} \texttt{R} package.}

\section{Case Study: PM$_{2.5}$ Air Pollution in the Greater London Region}\label{sec:application}

\subsection{Basis dimension selection}\label{sec:basissel}
We conducted a sensitivity analysis using a leave-one-site-out validation approach to determine the optimal number of dimensions, $m$, for the basis functions that model the temporal trends of the scale parameters $\sigma_{u_y,i},\sigma_{u_x,i}$.
These parameters define the matrices $\Phi_i$ and $\Psi_i$; see \eqref{eq:scales}.
The analysis evaluated changes in root mean square error (RMSE), mean absolute error (MAE), and the mean width of 95\% predictive intervals (mean PI) for values of $m$ ranging from 5 to 150. 
Results averaged across all locations and presented in Figure~\ref{fig:appendix_basis} of the Supplementary Material show minimal improvements in RMSE, MAE, and mean PI width beyond $m=60$.
\rev{Since the dominant contribution to the latent dimension comes from the basis-expansion coefficients, their total number is of order $8mn$. This count arises from four coefficient blocks, $\alpha,\beta,c,d$, each of size $m\times n$, repeated for the two data sources. Clearly, higher values of $m$ incur substantial computational costs. Therefore,}
we set $m=60$ as it represents the smallest dimension at which further improvements are negligible, balancing accuracy and computational efficiency.

\subsection{Priors for data hyperparameters}\label{sec:priors_data}
As mentioned in~\ref{sec:modelframework}, we assume the shape parameters $\xi_y$, $\xi_x$ to be constant over space and time to ensure identifiability and reduce computational complexity. 
Following~\cite{castro-camilo_practical_2022}, we impose \emph{a priori} restrictions over $\xi_y$, $\xi_x$ to ensure the existence of first and second moments and the regularity of MLEs.
In the spirit of PC priors {\citep{simpson_penalising_2017}}, we aim for priors that promote parsimony by encouraging simpler models. 
In the context of GPDs, the simplest model is obtained when the shape parameter equals zero.
To encapsulate these requirements, we define a scaled Laplace prior for $\xi\in\{\xi_y,\xi_x\}$ with location $\mu \in \mathbb{R}$ and scale $b > 0$:
\begin{equation}\label{eq:DF_laplace_prior}
    \text{Laplace}_{\xi}(\mu,b)   = 
    \begin{cases}
        \frac{1}{b}\left(\frac{\text{exp}\left(\frac{-|\xi-\mu|}{b}\right)}{2-\text{exp}\left(\frac{-0.5-\mu}{b}\right)-\text{exp}\left(\frac{\mu-0.5}{b}\right)}\right) & \quad \text{for } -0.5 < \xi < 0.5,\\
        0 & \quad \text{otherwise. }
    \end{cases}
\end{equation}
{Setting $\mu_x=\mu_y = 0$ and choosing a reasonably small $b$ (in our application, $b_x= b_y = 0.05$)  ensures that this prior concentrates most of its mass around zero. 
This effectively penalises deviations from zero while satisfying the conditions for regularity of MLEs ($\xi_y, \xi_x > -0.5$) and the existence of first and second moments ($\xi_y, \xi_x < 0.5$).}

\subsection{Priors for logistic regression parameters and hyperparameters}\label{sec:priors_log}
A sensitivity analysis was conducted to assess the impact of prior specifications on {$\boldsymbol\lambda_{i} = (\lambda_{i0}, \lambda_{i1}, \lambda_{i2}, \lambda_{i3})^\top$.}
The results were consistent across all locations.
As an illustrative example, Figure~\ref{fig:appendix_lambdas} of the Supplementary Material shows findings for site L, demonstrating that all components of {$\boldsymbol\lambda_{i}$} remain robust to prior specification.
Consequently, we assigned uninformative prior values to all components, specifically setting $\mu_{\lambda_l} = 0$ and $\sigma_{\lambda_l}^2=1$, $l=0,1,2,3$.

\subsection{Priors for other parameters and hyperparameters}\label{sec:priors_other}
Following~\cite{wilkie_data_2015}, the exponential decay parameters $\phi_\alpha$ and $\phi_\beta$ were treated as fixed within the hierarchical model~\eqref{eq:DF_dGPD}.
However, in the absence of prior knowledge regarding appropriate values for these parameters, an estimation procedure was required. To address this, we adopted a frequentist approach to estimate them externally before incorporating them as fixed values within the hierarchical model.
Specifically, we fitted a simple regression model of the form $\boldsymbol{y}^{*}_i = a_i + b_i\boldsymbol{x}^{*}_i$ independently for each location $i = 1,\ldots,n$. 
Exponential variograms were subsequently fitted to the estimated regression coefficients, yielding decay rates $\hat\phi_\alpha = \hat\phi_\beta = 1.6$, which were held constant throughout model fitting. 
This pragmatic approach captures key features of the observed data through a straightforward linear regression while maintaining computational efficiency.

For the remainder hyperparameters (listed in the bottom of Figure~\ref{fig:DF_FlowChart}), we choose the following values: $a_\alpha = 2$, $b_\alpha=1$, $a_\beta=2$, $b_\beta=1$, $a_c=2$, $b_c=1$, $\mathbf{\mu}_d =\mathbf{0} $, and $\Sigma_d$ is a multiple of $\mathbf{I}_m$, as suggested in \citet{wilkie_nonparametric_2019}.

\subsection{MCMC convergence}
To evaluate model convergence, two MCMC chains were initialised with randomised starting parameter values. 
Each chain ran for 3 million iterations, with the first 1 million discarded as burn-in. 
Convergence was robust, as indicated by trace and density plots (not shown). 
The Gelman-Rubin convergence diagnostic yielded point estimates of 1, confirming strong convergence across all parameters. 
Notably, convergence was achieved with as few as 500,000 iterations and a burn-in of 100,000 samples, demonstrating computational efficiency.

\subsection{Goodness of fit}
\begin{figure}
    \centering
    \includegraphics[scale=0.47]{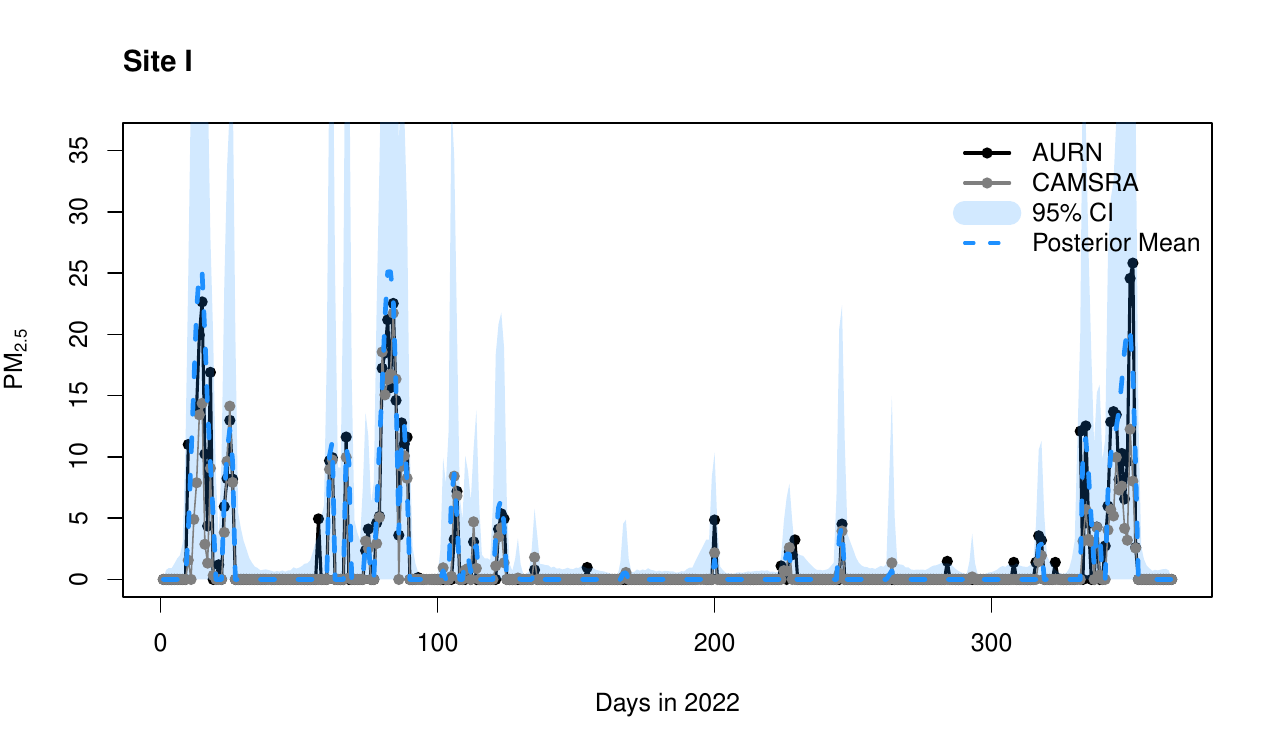}
     \includegraphics[scale=0.47]{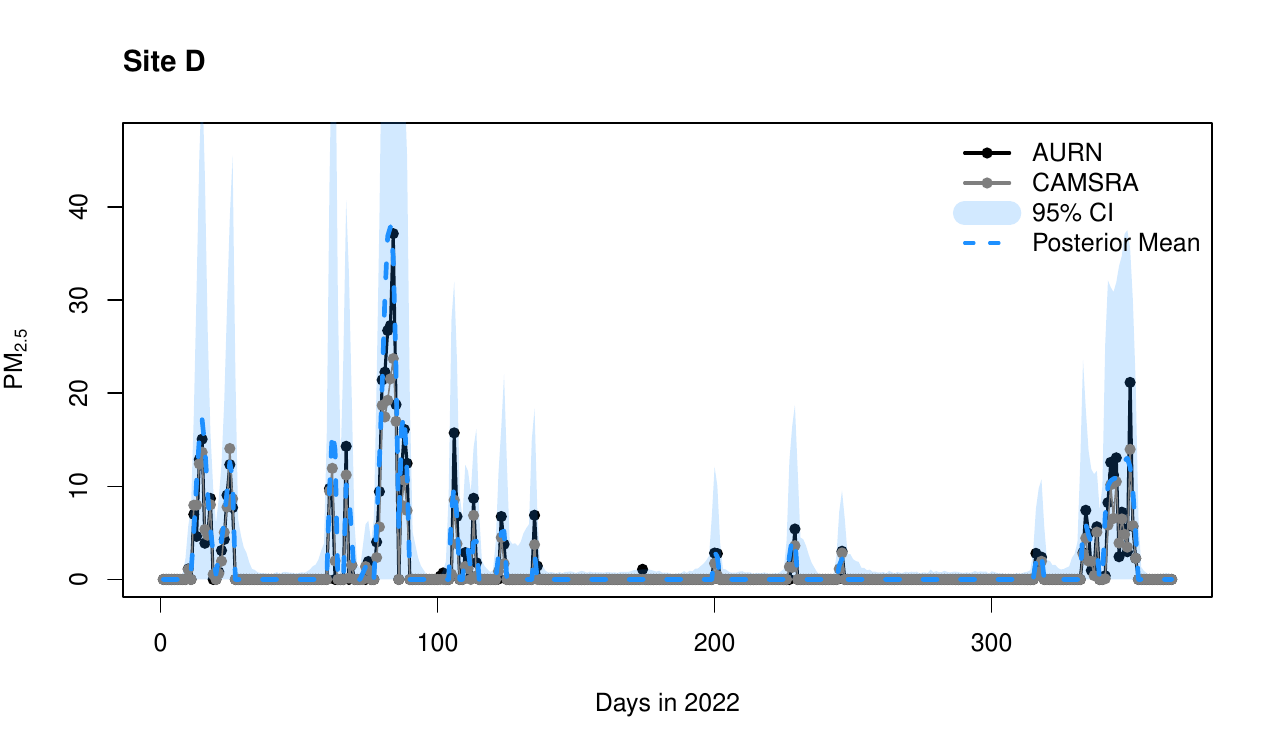}
    \includegraphics[scale=0.47]{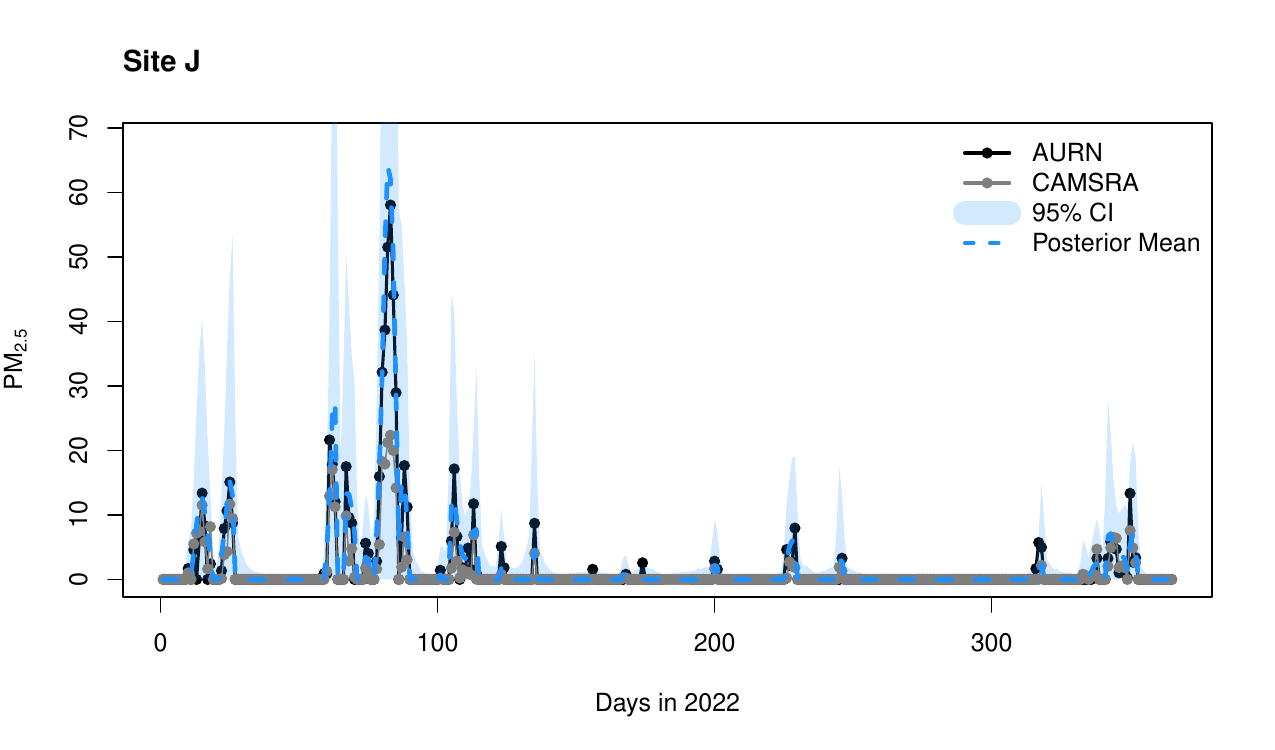}
     \caption{\PMp{} measurements from sites I (West), D (central London) and J (East) are displayed in black for the AURN data, in grey for the CAMSRA values, and in blue for our ExDF model. 95\% pointwise credible intervals are provided for the ExDF model.}
    \label{fig:gof}
\end{figure}

Figures~\ref{fig:gof} illustrate the overall goodness of fit of our model at three specific locations: site I to the west, site D near central London, and site J to the east. 
The plots display posterior means and 95\% credible intervals, as well as observations from the AURN and CAMSRA datasets for comparison.
The posterior means closely align with the AURN data, effectively capturing both threshold exceedances ($\boldsymbol{y}_i>0$) and censored non-exceedances ($\boldsymbol{y}_i=0$). 
The 95\% credible intervals exhibit high coverage, missing only two observations on average, corresponding to a coverage probability of 0.978, although the intervals are relatively wide. 
For instance, at the timestamp $83$ (the yearly maximum), the upper bounds are 120 for site D and 85 for site I.
Posterior density plots for the shape parameters $\xi_y$ and $\xi_x$ (Figure~\ref{fig:appendix_density_xis} in the Supplementary Material) show means of -0.10 (-0.216, 0.05, 95\% CI) and -0.22 (-0.34, 0.02, 95\% CI), respectively. 
Although pointwise posterior estimates show slight differences, both datasets suggest unbounded tails when uncertainty is considered.
Furthermore, the ExDF model more accurately captures AURN threshold exceedances than the CAMSRA data, particularly \rev{in the upper tail.}
\rev{This is supported by Figure~\ref{fig:scatterplots} in the Supplementary Material, which presents scatterplots comparing AURN exceedances with the corresponding CAMSRA values and the fitted exceedances (posterior means) from the ExDF and GaussDF models across all 12 monitoring sites.}

\subsection{Model validation} A leave-one-site-out cross-validation (LOSO-CV) approach was used on the twelve AURN sites to assess the model's predictive performance. 
In each iteration of this LOSO-CV, a site was removed from the dataset used to fit the model. 
The fitted model was then used to provide a prediction for the site that was initially removed. 
This procedure was repeated for all 12 AURN sites. 
{Predictions were compared against observed exceedances in the AURN data and evaluated using continuous rank probability score (CRPS), root mean square error (RMSE), and mean absolute error (MAE).
For comparison, predictions from the ExDF model were assessed alongside those from the Gaussian model of \citet{wilkie_nonparametric_2019} fitted to the whole \rev{log-transformed PM$_{2.5}$ data and subsequently back-transformed} (GausDF) and the CAMSRA data at each site. 
Table~\ref{tab:DF_LOOCV} highlights two key findings. 
First, both data fusion models provide more accurate representations of extreme values observed \emph{in situ} than \rev{the gridded reanalysis data alone}. 
Specifically, ExDF and GausDF outperform CAMSRA at 10 and 7 sites, respectively, though CAMSRA performs better at sites C and G based on RMSE and MAE. 
Second, the ExDF model consistently outperforms the Gaussian model at every site, as reflected in lower RMSE, MAE, and CRPS values, demonstrating that specialised extreme-value approaches are better suited for capturing extremes than Gaussian models.
}

\begin{table}
\centering
\caption{LOSO-CV results comparing the predictions of the GausDF and ExDF models with the CAMSRA data. The table reports RMSE, MAE, and CRPS values for sites A through L across the three data sources, with boldface indicating the lowest value for each metric at a given site.}
\centering

\begin{tabular}{lrrrrrrrr}
& & & & & & & & \\
\toprule
                                               & \multicolumn{3}{c}{\textbf{RMSE}}                                                                                    & \multicolumn{3}{c}{\textbf{MAE}}                                                             & \multicolumn{2}{c}{\textbf{CRPS}} \\ \hline
\multicolumn{1}{l|}{Site}                      & \cellcolor[HTML]{FFFFFF}GausDF & ExDF          & \multicolumn{1}{r|}{CAMSRA}                                  & GausDF & ExDF          & \multicolumn{1}{r|}{CAMSRA}                                  & GausDF  & ExDF           \\ \hline
\rowcolor[HTML]{EFEFEF} 
\multicolumn{1}{l|}{\cellcolor[HTML]{EFEFEF}A} & 3.12                           & \textbf{1.68} & \multicolumn{1}{r|}{\cellcolor[HTML]{EFEFEF}7.57}          & 2.79   & \textbf{0.94} & \multicolumn{1}{r|}{\cellcolor[HTML]{EFEFEF}6.44}          & 0.52    & \textbf{0.35}  \\
\multicolumn{1}{l|}{B}                         & 2.25                           & \textbf{1.20} & \multicolumn{1}{r|}{5.01}                                  & 2.12   & \textbf{0.83} & \multicolumn{1}{r|}{4.03}                                  & 0.52    & \textbf{0.36}  \\
\rowcolor[HTML]{EFEFEF} 
\multicolumn{1}{l|}{\cellcolor[HTML]{EFEFEF}C} & 2.21                           & 1.10          & \multicolumn{1}{r|}{\cellcolor[HTML]{EFEFEF}\textbf{0.62}} & 1.66   & 0.68          & \multicolumn{1}{r|}{\cellcolor[HTML]{EFEFEF}\textbf{0.50}} & 0.52    & \textbf{0.36}  \\
\multicolumn{1}{l|}{D}                         & 1.71                           & \textbf{1.21} & \multicolumn{1}{r|}{1.94}                                  & 1.22   & \textbf{0.77} & \multicolumn{1}{r|}{0.99}                                  & 0.52    & \textbf{0.38}  \\
\rowcolor[HTML]{EFEFEF} 
\multicolumn{1}{l|}{\cellcolor[HTML]{EFEFEF}E} & 5.30                           & \textbf{2.20} & \multicolumn{1}{r|}{\cellcolor[HTML]{EFEFEF}5.89}          & 5.12   & \textbf{1.64} & \multicolumn{1}{r|}{\cellcolor[HTML]{EFEFEF}5.23}          & 0.47    & \textbf{0.29}  \\
\multicolumn{1}{l|}{F}                         & 3.89                           & \textbf{1.66} & \multicolumn{1}{r|}{3.12}                                  & 3.75   & \textbf{1.01} & \multicolumn{1}{r|}{2.00}                                  & 0.48    & \textbf{0.34}  \\
\rowcolor[HTML]{EFEFEF} 
\multicolumn{1}{l|}{\cellcolor[HTML]{EFEFEF}G} & 3.53                           & 3.09          & \multicolumn{1}{r|}{\cellcolor[HTML]{EFEFEF}\textbf{2.29}} & 3.35   & 2.38          & \multicolumn{1}{r|}{\cellcolor[HTML]{EFEFEF}\textbf{1.57}} & 0.49    & \textbf{0.19}  \\
\multicolumn{1}{l|}{H}                         & 2.92                           & \textbf{2.38} & \multicolumn{1}{r|}{2.60}                                  & 2.56   & \textbf{1.66} & \multicolumn{1}{r|}{1.75}                                  & 0.50    & \textbf{0.18}  \\
\rowcolor[HTML]{EFEFEF} 
\multicolumn{1}{l|}{\cellcolor[HTML]{EFEFEF}I} & 3.80                           & \textbf{2.66} & \multicolumn{1}{r|}{\cellcolor[HTML]{EFEFEF}2.83}          & 3.66   & \textbf{2.45} & \multicolumn{1}{r|}{\cellcolor[HTML]{EFEFEF}2.57}          & 0.50    & \textbf{0.15}  \\
\multicolumn{1}{l|}{J}                         & 6.67                           & \textbf{5.50} & \multicolumn{1}{r|}{10.29}                                 & 5.90   & \textbf{3.42} & \multicolumn{1}{r|}{8.09}                                  & 0.48    & \textbf{0.18}  \\
\rowcolor[HTML]{EFEFEF} 
\multicolumn{1}{l|}{\cellcolor[HTML]{EFEFEF}K} & 3.33                           & \textbf{3.11} & \multicolumn{1}{r|}{\cellcolor[HTML]{EFEFEF}6.78}          & 2.30   & \textbf{2.12} & \multicolumn{1}{r|}{\cellcolor[HTML]{EFEFEF}4.59}          & 0.51    & \textbf{0.18}  \\
\multicolumn{1}{l|}{L}                         & 5.30                           & \textbf{2.10} & \multicolumn{1}{r|}{8.49}                                  & 5.03   & \textbf{0.98} & \multicolumn{1}{r|}{7.31}                                  & 0.49    & \textbf{0.34}  \\
\bottomrule
\hline
\end{tabular}
\label{tab:DF_LOOCV}
\end{table}

\begin{figure}
    \centering
     \includegraphics[scale=0.55]{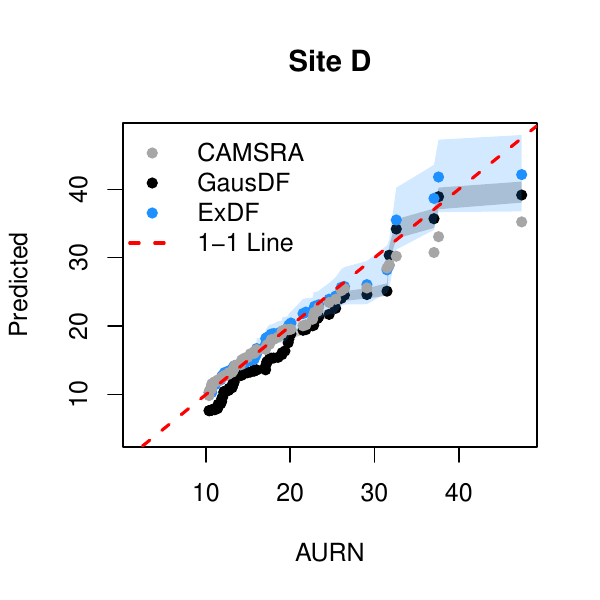}
    \includegraphics[scale=0.55]{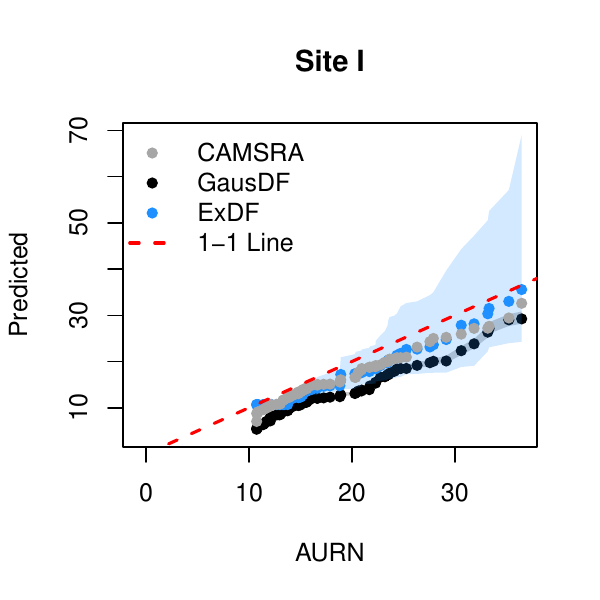}
    \includegraphics[scale=0.55]{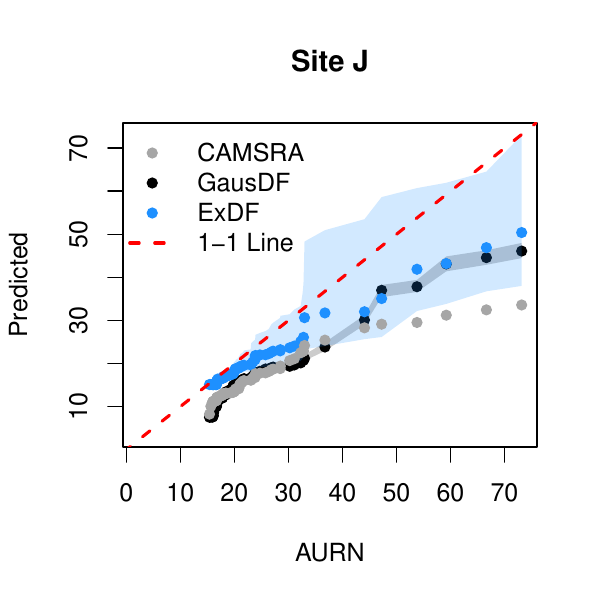}
    \caption{Q-Q plots of LOSO-CV results for sites I (West), D (central London) and J (East) are displayed in black for the GausDF model \citep{wilkie_nonparametric_2019}, in grey for the CAMSRA values, and in blue for our ExDF model. {Monte Carlo-based} Pointwise 95\% confidence intervals, {derived from 1,000 posterior samples,} are shown for both the ExDF and GausDF models.}
    \label{fig:qqplots}
\end{figure}

{The Q-Q plots in Figures \ref{fig:qqplots} for sites I, D and J provide a visual representation of the validation results summarised in Table \ref{tab:DF_LOOCV}.}
{The ExDF model demonstrates superior performance in capturing observed values at site D compared to both the CAMSRA data and the GausDF model.
Threshold exceedances at site I are also well captured by the ExDF model, although CI are wider for larger values of the AURN dataset.
For site J, we can see that the largest threshold exceedances are consistently underestimated, though this trend is captured within the ExDF confidence intervals.}
\rev{This poorer fit may reflect the fact that the gridded background field departs more strongly from the local station behaviour there than at the other sites. In addition, because the model uses global shape parameters, some of the residual underestimation at the largest values may be due to site-specific tail behaviour that is not fully captured under this parsimonious specification.}

{To assess the predictive performance of the logistic regression component, we computed additional validation and classification metrics, evaluating its ability to accurately predict and classify threshold exceedances. 
Table~\ref{tab:appendix_class_logreg} in the Supplementary Material presents these classification metrics for the logistic regression component of the ExDF model, applied to fitted sites A--K when predicting at site L. 
The table also includes a comparison with the CAMSRA data. 
While classification performance varies across sites, overall, the model demonstrates high accuracy across all metrics.}

{Overall, our results indicate that the ExDF model outperforms both the CAMSRA \rev{data} and \rev{the} GausDF model in capturing the true threshold exceedances observed at the \emph{in situ} stations. In some cases, the ExDF model exhibits greater uncertainty, which is expected given that it is fitted to censored threshold exceedances, whereas the GausDF approach is fitted to the full dataset. A notable limitation of the ExDF model is that, at locations where \rev{the gridded reanalysis field} poorly \rev{represents} actual conditions, the improvements achieved by data fusion models over the CAMSRA data are minimal and may be deemed negligible.}

\subsection{Spatial predictions} 
\begin{figure}
    \centering
    \includegraphics[scale=0.5]{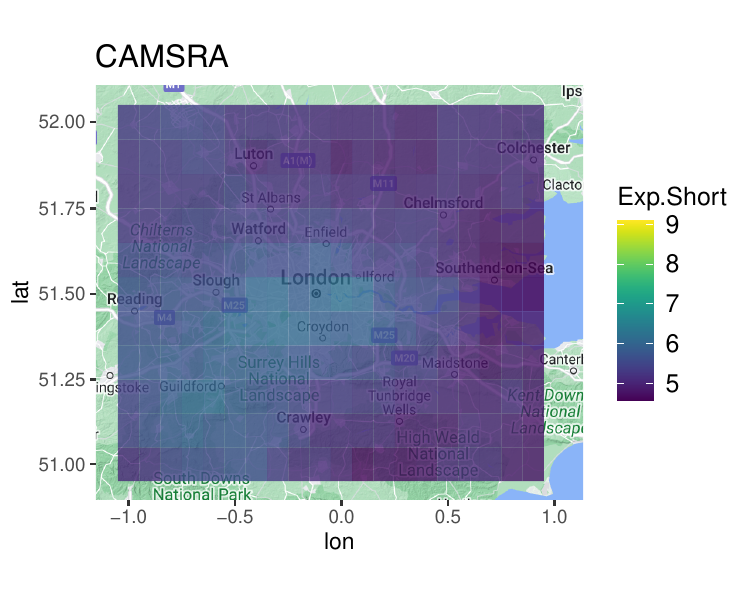}
    \includegraphics[scale=0.5]{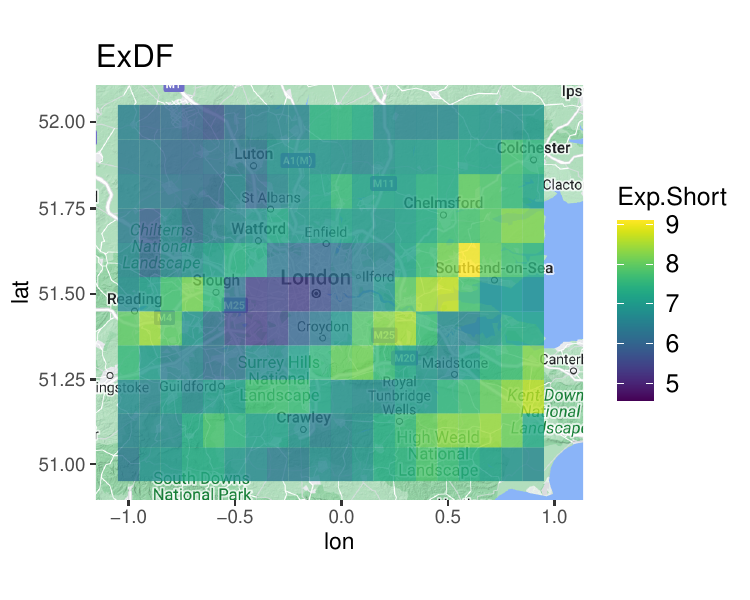}
    \includegraphics[scale=0.485]{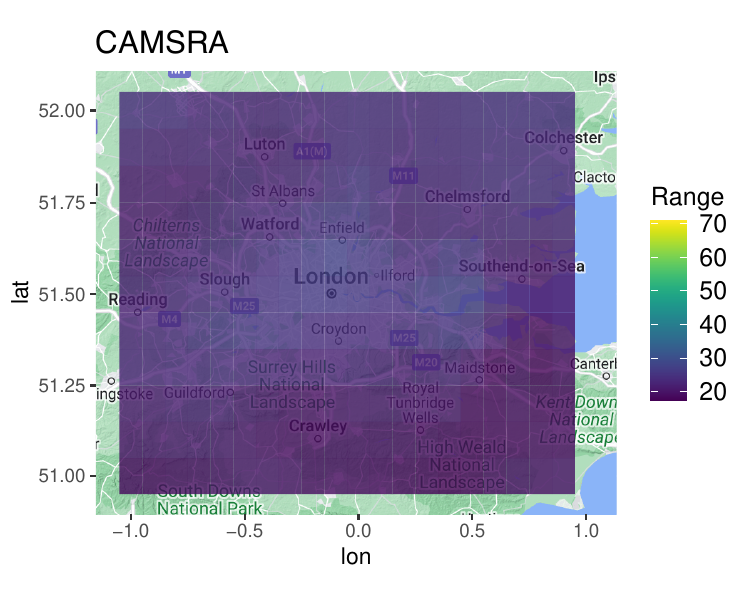}
    \includegraphics[scale=0.485]{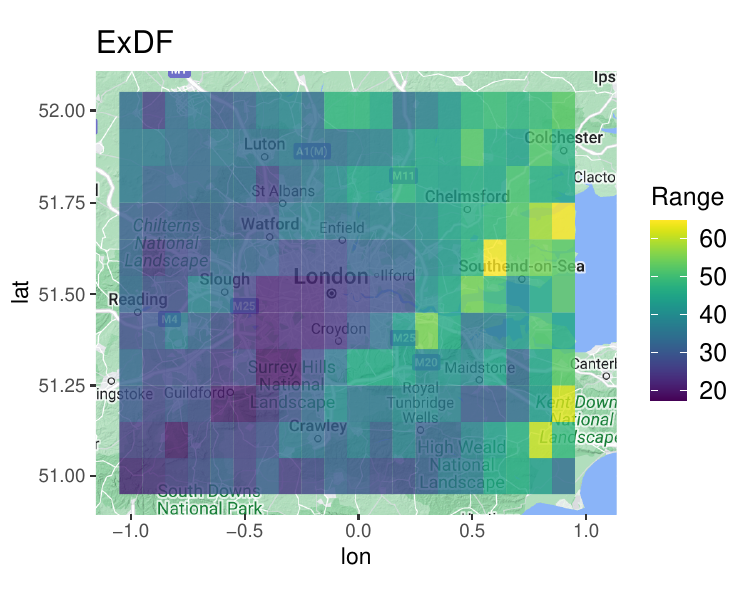}
    \caption[Maps of \PMp{} expected shortfall for exceedances from the CAMSRA data and the ExDF model.]{\emph{Top}: Map of \PMp{} expected shortfall from the CAMSRA data and the ExDF model. \emph{Bottom:} Map of the range of exceedances for the CAMSRA data and the ExDF model. }
    \label{fig:DF_mapity_maps}
\end{figure}

Predictions of \PMp{} concentrations across the Greater London region were generated using the ExDF model and compared against those from CAMSRA. 
Figure \ref{fig:DF_mapity_maps} illustrates the expected shortfall (i.e., the average of threshold exceedances) and the range of threshold exceedances (i.e., difference between maximum and minimum value), for both the CAMSRA data and ExDF predictions. 
Together, these two statistics capture the location and spread of threshold exceedances.
The expected shortfall estimates in the top row of Figure \ref{fig:DF_mapity_maps} show that the CAMSRA data exhibit minimal variability, with smooth spatial patterns. 
The largest expected shortfall value of \SI{6.2}{\micro\gram}/m$^3$ is observed in central London, followed by the southwestern region. In contrast, the ExDF model reveals different spatial patterns, with the largest shortfall value of \SI{9.07}{\micro\gram}/m$^3$ located on the east coast. Additionally, diagonal patterns in both ExDF plots align with the directions of high-frequency wind gusts in the region.

\begin{figure}
    \centering
    \includegraphics[scale=0.7]{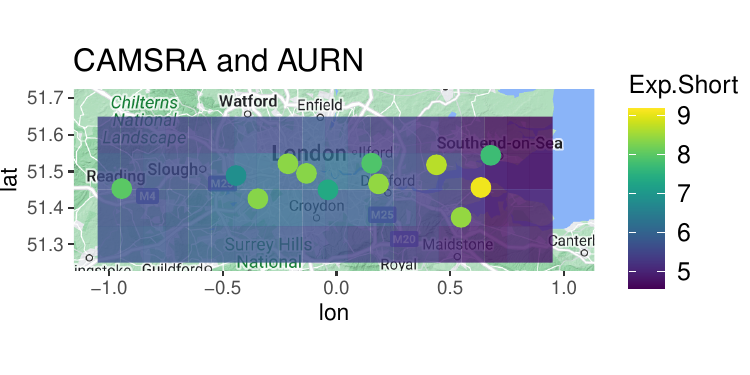}
    \includegraphics[scale=0.7]{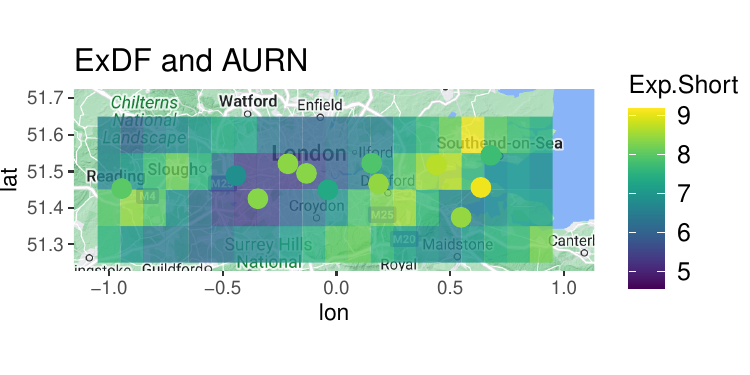}
    \caption[Cropped maps of \PMp{} expected shortfall for exceedances from the CAMSRA data and the ExDF model with AURN observations stations.]{Cropped map of the Greater London area showing \PMp{} expected shortfall from the CAMSRA (top) and ExDF (bottom) data along with the empirical shortfall computed at each of the 12 AURN observation stations. }
    \label{fig:DF_mapity_maps2}
\end{figure}

{Figure \ref{fig:DF_mapity_maps2} compares the expected shortfall predicted by the CAMSRA \rev{data} and \rev{the} ExDF model against observations from the AURN stations. The CAMSRA data shows overly smooth threshold exceedances, especially near the coast, while the ExDF model provides a more accurate representation, particularly at sites outside the city centre. 
\rev{In central London, however, the ExDF surface tends to slightly {underestimate} the observed expected shortfall at some stations, likely reflecting spatial smoothing inherited from the gridded background field and latent structure, which can damp highly localised urban peaks.}
The higher \PMp{} values along the coast are consistent with findings from \citet{yang_maritime_2023}, which attributed the increased concentrations to emissions from maritime transport and other related sources, as well as salt content from the sea.}
{Overall, the maps emphasise the differences between threshold exceedances in the CAMSRA data and the ExDF predictions---highlighting the value of data fusion in applications that focus on extreme values.}

\section{Discussion and Conclusions}\label{sec:DF_DandC} 
{Extreme values of \PMp{} air pollution pose significant risks to public health, particularly in urban regions like Greater London. 
Accurate, local-scale measurement of \PMp{} is crucial for managing public exposure and mitigating harm. 
However, the limited spatial coverage of observation networks and the over-smoothing of \rev{gridded reanalysis products} challenge the accuracy of pollution assessments. 
Our ExDF model addresses these challenges by fusing datasets to produce a spatially complete and temporally detailed representation of \PMp{} concentrations.}

{The ExDF model extends the data fusion framework of \citet{wilkie_nonparametric_2019} to integrate \emph{in situ} and \rev{gridded CAMSRA background data} for \PMp{} concentrations. 
Specifically, it links the two datasets through the scale parameter of the \dGPD{} by employing a flexible regression framework with parameters { that vary in time and space}. 
Unlike existing models that use EVT for data fusion, the ExDF model retains information about the time of a threshold exceedance by {censoring and }accounting for non-threshold exceedances. 
Despite its strengths, the model's reliance on threshold exceedances in the \rev{gridded reanalysis field} presents a limitation, as it struggles to predict exceedances in \emph{in situ} data when they are absent in the \rev{gridded data}. Reassessing and incorporating additional covariates, such as those identified by \citet{jin_machine_2022} (e.g., dew point temperature, humidity, vegetation indices, and drought indices), could enhance the predictive performance.}

{Our ExDF model offers a significant advantage by preserving the temporal structure of exceedances, which is a desirable feature for applications requiring detailed spatio-temporal analyses. 
However, the framework may introduce discontinuities near the threshold, suggesting that future work could explore continuous models capable of jointly capturing extreme and non-extreme observations, such as those based on the extended generalised Pareto distribution \citep{naveau_modeling_2016}. 
{Additionally, our approach strikes a balance between parsimony and practicality. For example, the assumption that $\xi_x$ and $\xi_y$ are constant and independent was made to simplify the number of estimated parameters.}
\rev{However, the behaviour at site J suggests that some residual lack of fit at the most extreme values may be attributable to site-level tail heterogeneity that is not represented when shape parameters are held constant across locations.
Therefore}, simple models that account for potential relationships between $\xi_x$ and $\xi_y$ could be considered. 
That said, increasing the number of parameters can result in inferential challenges, potentially jeopardising the identifiability of the linear relationship between scale parameters (namely, parameters $c_i$ and $d_i$), which already link the two datasets. }
{Spatial structure is incorporated through an exponential decay model with parameters estimated \textit{a priori} from variograms of proxy parameters. While effective in this study, this approach assumes second-order spatial stationarity, which may not generalise to all datasets.}

\rev{The proposed framework is not specific to \PMp{} and could, in principle, be adapted to other environmental variables for which one wishes to link coarse-scale products and site-level extremes, such as rainfall or temperature. That said, a proper worked-out extension would require additional case-specific modelling choices rather than a direct substitution of the data source. For rainfall, for example, one would need to reconsider the threshold selection in light of zero inflation and event intermittency, decide whether the temporal basis structure should reflect seasonality and storm-type heterogeneity more explicitly, and potentially replace the current collocation step with one better suited to convective systems and sub-grid spatial variability. One might also need to revisit the tail model itself if accumulation duration or event dependence were central to the application.}

\rev{A longer study period could also be informative and, in principle, could improve estimation by increasing the number of exceedances available for inference. Our choice to focus on 2022 was deliberate. The aim of this paper is to introduce and demonstrate a workable and adaptable modelling framework, rather than to fully optimise predictive performance over the longest possible record. Restricting attention to a single recent year allows us to present the methodology in a comparatively simple setting, without introducing additional temporal structure that would likely be needed over a longer multi-year period, such as longer-term trends, interannual variability, or more complex seasonal effects. Extending the analysis to multiple years is certainly feasible within the proposed framework, but would require additional modelling choices and computational work.}

\rev{A similar point applies more generally to the choice of collocation rule. Because CAMSRA is already a coarse, spatially averaged background product, using neighbouring cells would mainly refine how that background information is supplied to the model rather than remove the need for local calibration. In the present work we therefore preferred the simpler nearest-centroid linkage and allowed the main adjustment from coarse-scale reanalysis data to site-level extremes to be learned through the hierarchical tail model.}

{Results from Section \ref{sec:application} demonstrate the model's effectiveness, with the posterior predictive mean providing reliable point predictions. 
Specifically, a leave-one-site-out cross-validation procedure demonstrated that the ExDF model outperforms both the GausDF \rev{model} and \rev{the} CAMSRA \rev{data} in capturing threshold exceedances for 10 out of 12 \emph{in situ} sites, highlighting its robustness in integrating diverse data sources. Nonetheless, its performance is limited in areas where \rev{the gridded reanalysis field} is biased or poorly represents the phenomenon.}

{Additional comparisons between the ExDF \rev{model} and \rev{the} CAMSRA \rev{data} reveal notable differences. 
For instance, maps of expected shortfall show that the CAMSRA data are spatially smooth with low variability, whereas the ExDF predictions exhibit greater variability and capture spatial patterns, such as higher \PMp{} concentrations near coastal areas. These differences underscore the ExDF model's ability to uncover spatial patterns and intensities of \PMp{} pollution that are less apparent in \rev{the gridded reanalysis data alone}.
As noted earlier, the performance of the ExDF model is highly dependent on the quality of \rev{the gridded data}. 
This highlights the importance of selecting reliable \rev{gridded products}, as is the case with any data fusion model.}

{Overall, the ExDF model provides a powerful framework for integrating \emph{in situ} observations and \rev{gridded CAMS reanalysis data}, offering comprehensive temporal records and revealing critical spatial patterns not apparent in single-source datasets. Future research should explore the inclusion of additional covariates, alternative models for the entire distribution, and methods to address potential biases in \rev{gridded background products}, further enhancing the utility and reliability of data fusion approaches for extreme-value analyses.}



\appendix
\section*{Appendix}
\begin{figure}
    \centering
    \includegraphics[scale=0.54]{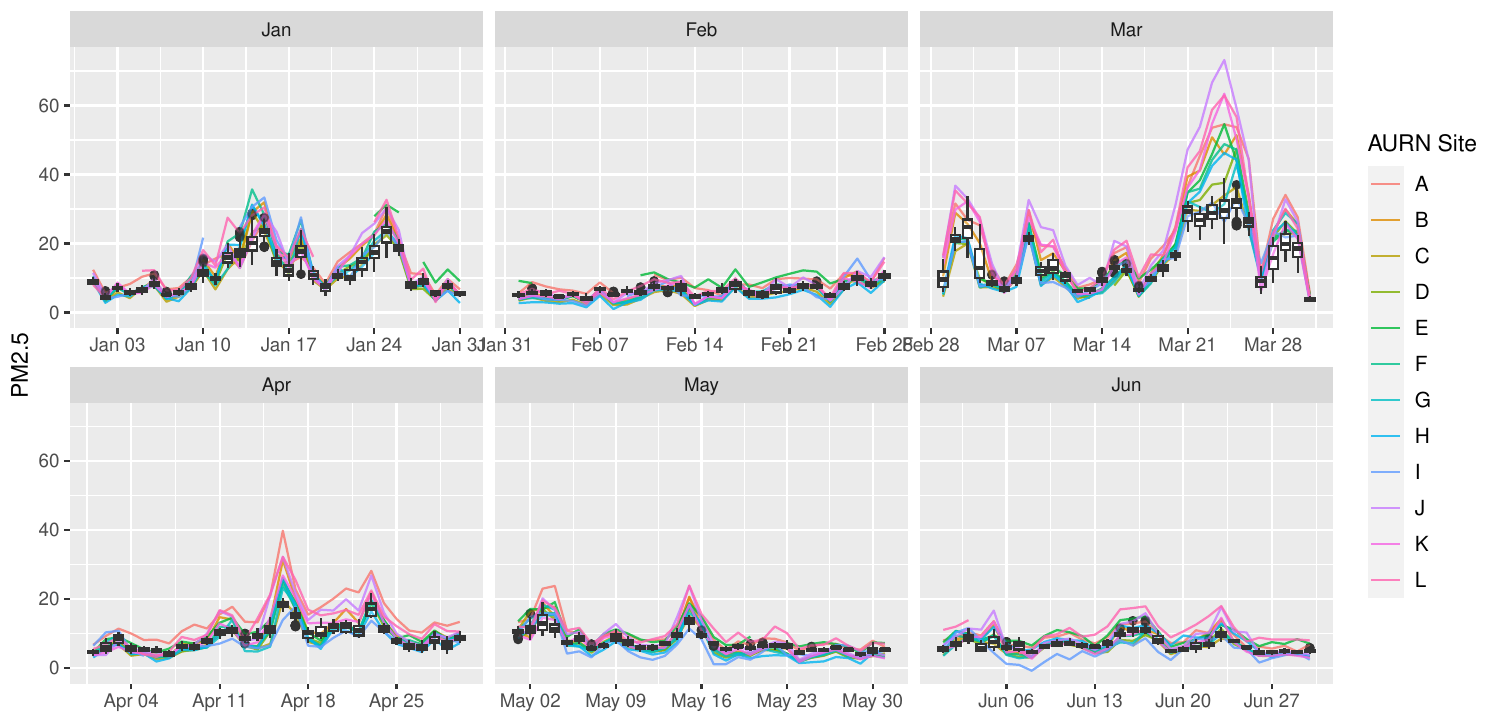}
    \includegraphics[scale=0.54]{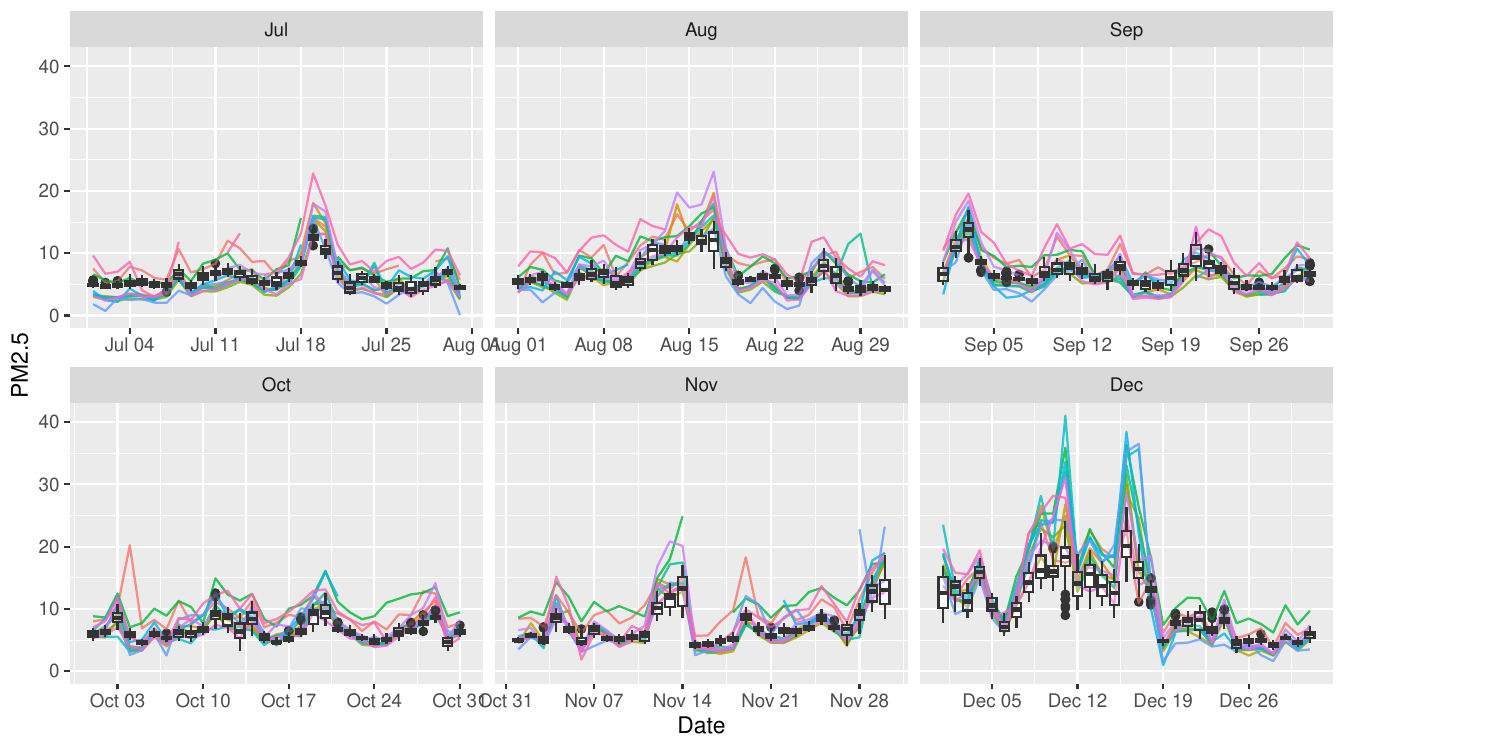}
    \caption{Boxplots for daily mean concentrations of \PMp{} for January to December 2022 as provided by the CAMSRA data. The daily concentrations of the 12 AURN sites in the area are given as lines. }
    \label{fig:appendix_NS_DF_CAMS_ts1}
\end{figure}

\rev{This supplement provides additional data visualisations and results supporting the case study on \PMp{}.}
Figure~\ref{fig:appendix_NS_DF_CAMS_ts1} illustrates the temporal evolution of AURN and CAMSRA data across all months of 2022. Boxplots summarise the distribution of daily mean concentrations from CAMSRA, while the corresponding observations at the 12 AURN monitoring sites are overlaid as time series.

\begin{figure}
    \centering
    \includegraphics[scale=0.65]{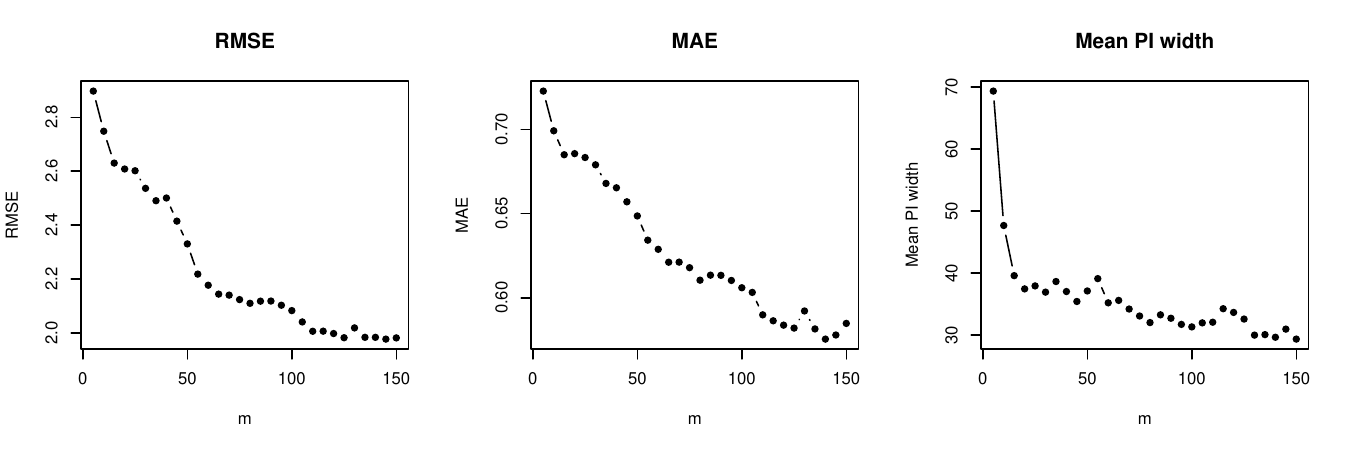}
    \caption{Sensitivity analysis using a \rev{leave-one-site-out} validation approach to determine the optimal $m$, which represents the number of dimensions for the basis functions that model the temporal trends of the GPD scale parameters. The figure shows RMSE, MAE and mean width of 95\% predictive intervals (mean PI) for $m=5,\ldots,150$ averaged across all locations. }
    \label{fig:appendix_basis}
\end{figure}

\begin{figure}
    \centering
    \includegraphics[scale=0.5]{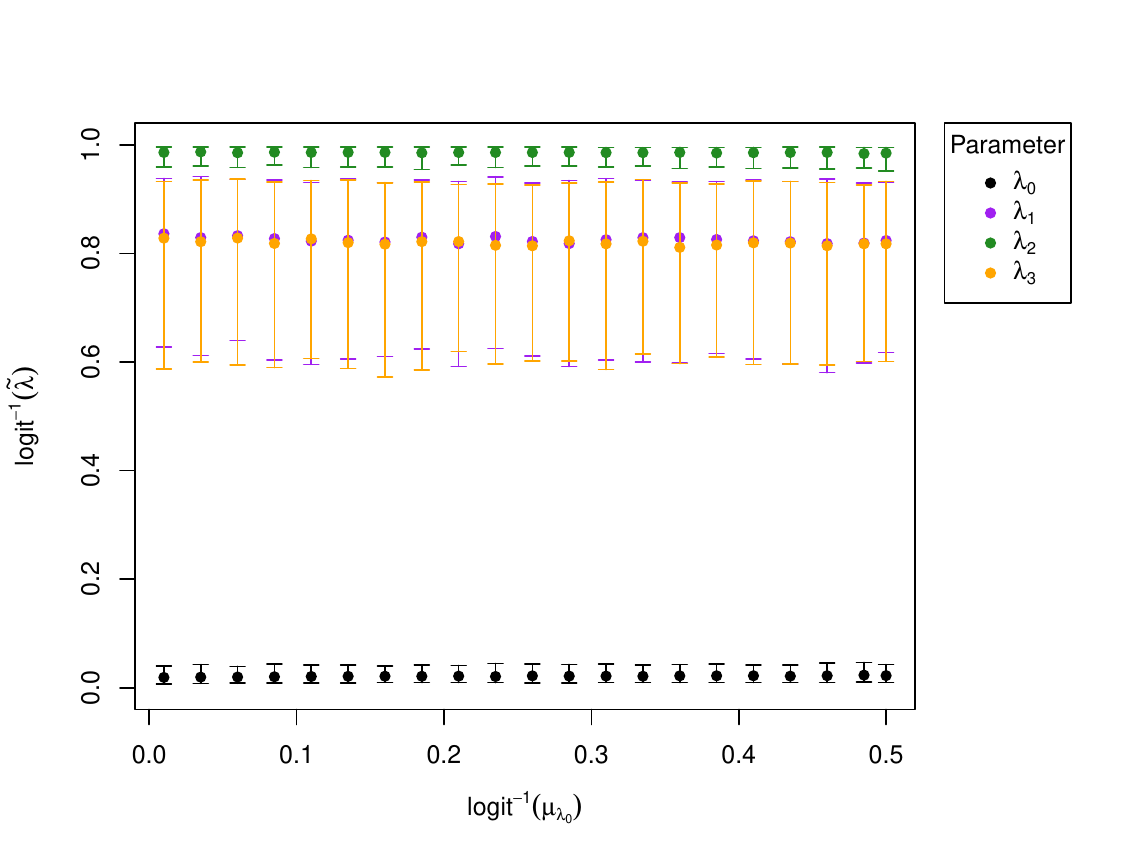} 
    \caption{Sensitivity analysis conducted for site L to evaluate the impact of prior specifications on $\boldsymbol\lambda_{i} = (\lambda_{i0}, \lambda_{i1}, \lambda_{i2}, \lambda_{i3})^\top$. The plot shows posterior mean estimates and 95\% credible intervals for the posterior samples of each component $\lambda_{i}^{(l)}$ against various values for $\mu_{\lambda_l}$, $l=0,1,2,3$. Similar studies were conducted for the remaining sites.}
    \label{fig:appendix_lambdas}
\end{figure}

\begin{figure}
    \centering
    \includegraphics[scale=0.88]{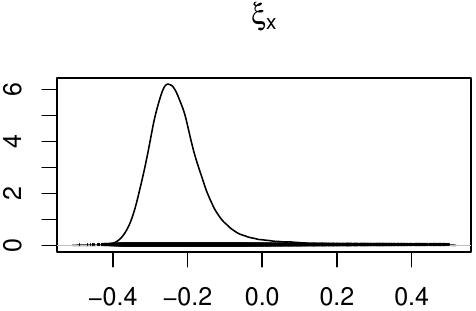}
    \includegraphics[scale=0.88]{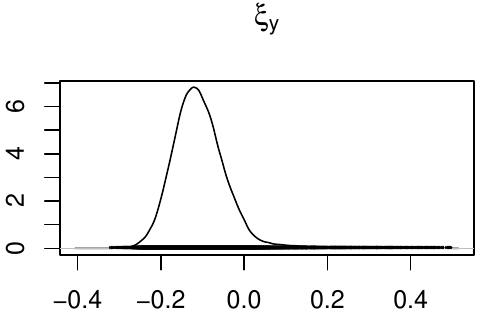}
    \caption{Posterior density plots for $\xi_y$ and $\xi_x$.}
    \label{fig:appendix_density_xis}
\end{figure}

Further results related to model fitting and performance are also provided. Figure~\ref{fig:appendix_basis} shows the root mean square error, mean absolute error, and mean width of 95\% predictive intervals as a function of the basis dimension $m$, informing the choice of basis dimension for modelling the temporal trends of the GPD scale parameters. Figure~\ref{fig:appendix_lambdas} presents posterior means and 95\% credible intervals for the coefficients $\boldsymbol{\lambda}_i$ at location L, based on predictions using data from locations A--K. Figure~\ref{fig:appendix_density_xis} displays posterior density estimates for the GPD shape parameters $\xi_y$ and $\xi_x$.

\begin{figure}
    \centering
    \includegraphics[width=\linewidth]{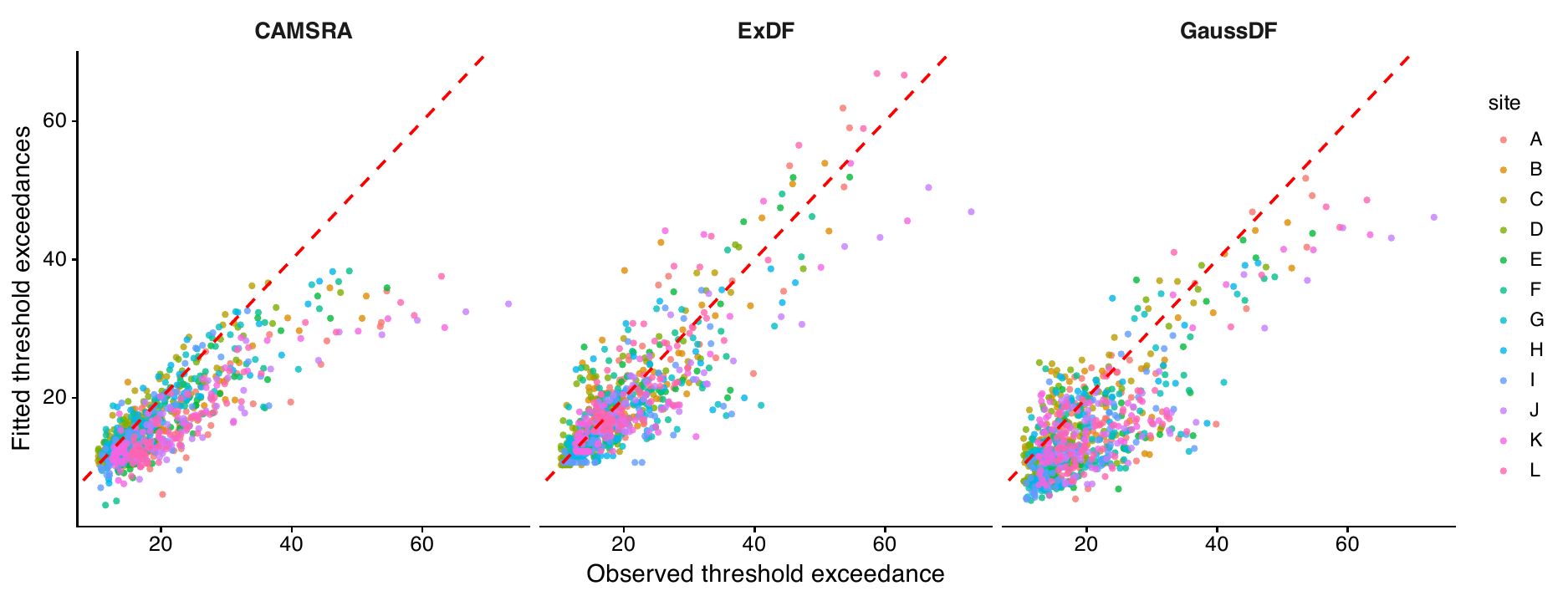}
    \caption{\rev{Observed threshold exceedances from AURN against the corresponding CAMSRA exceedances and the fitted exceedances (posterior means) from ExDF and GaussDF, for all 12 monitoring sites.}}
    \label{fig:scatterplots}
\end{figure}

Table~\ref{table:classification} reports standard classification metrics for the logistic regression component of the ExDF model across sites A--K. 
\rev{Finally, Figure~\ref{fig:scatterplots} compares observed AURN exceedances with corresponding CAMSRA exceedances and fitted exceedances (posterior means) from the ExDF and GaussDF models across all 12 monitoring sites.}

\begin{table}
\centering
\label{table:classification}
\caption{Classification metrics for the logistic regression component of the ExDF model for fitted sites A to K when predicting over site L. Table includes a comparison with the CAMSRA data.}
\centering
\begin{tabular}{rrrrrr}
& & & & & \\
\toprule
\rowcolor[HTML]{FFFFFF} 
\textbf{Site} & \textbf{Accuracy} & \textbf{Precision} & \textbf{Recall} & \textbf{Specificity} & \textbf{F1 Score} \\ \hline
\rowcolor[HTML]{EFEFEF} 
A    & 0.94     & 0.84      & 0.89   & 0.96        & 0.86      \\
B    & 0.96     & 0.87      & 0.95   & 0.97        & 0.91      \\
\rowcolor[HTML]{EFEFEF} 
C    & 0.96     & 0.87      & 0.92   & 0.97        & 0.89      \\
D    & 0.97     & 0.90      & 0.95   & 0.97        & 0.92      \\
\rowcolor[HTML]{EFEFEF} 
E    & 0.91     & 0.79      & 0.76   & 0.95        & 0.77      \\
F    & 0.94     & 0.83      & 0.89   & 0.95        & 0.86      \\
\rowcolor[HTML]{EFEFEF} 
G    & 0.95     & 0.85      & 0.92   & 0.96        & 0.88      \\
H    & 0.92     & 0.77      & 0.85   & 0.93        & 0.81      \\
\rowcolor[HTML]{EFEFEF} 
I    & 0.90     & 0.75      & 0.79   & 0.93        & 0.77      \\
J    & 0.95     & 0.84      & 0.92   & 0.96        & 0.88      \\
\rowcolor[HTML]{EFEFEF} 
K    & 0.95     & 0.86      & 0.92   & 0.96        & 0.89      \\
CAMSRA & 0.95     & 0.87      & 0.87   & 0.97        & 0.87      \\ 
\bottomrule
\end{tabular}
\label{tab:appendix_class_logreg}
\end{table}

\newpage
\bibliographystyle{CUP} 
\bibliography{bibliography}

\end{document}